\documentclass[twocolumn]{aa}

\def\Crat{\mbox{$^{12}$C/$^{13}$C}}

\usepackage{graphicx}
\usepackage{txfonts}
\begin{document}
   \title{Lithium and zirconium abundances 
     in massive Galactic O-rich AGB stars\thanks{Based
on observations obtained at the 4.2 m William Herschel Telescope operated on the island of
La Palma by the Isaac Newton Group in the Spanish Observatorio del Roque de Los
Muchachos of the Instituto de Astrofisica de Canarias. Also, based on
observations obtained with the ESO 3.6 m telescope at La Silla Observatory
(Chile)}}

   \subtitle{}

   \author{D.A. Garc\'{\i}a-Hern\'andez
          \inst{1} \and P. Garc\'{\i}a-Lario\inst{1,2} \and 
	  B. Plez\inst{3} \and A. Manchado\inst{4,5} \and F. D'Antona\inst{6} 
	 \and J. Lub\inst{7} \and H. Habing\inst{7} 
          }

   \offprints{D.A. Garc\'{\i}a-Hern\'andez}

   \institute{ISO Data Centre, Research and 
Scientific Support Department of ESA. European Space Astronomy Centre (ESAC), Villafranca del 
Castillo. P.O. Box  50727. E-28080 Madrid. Spain \email{Anibal.Garcia@sciops.esa.int} 
\and Herschel Science Centre, Research and Scientific Support Department of ESA. European Space Astronomy Centre (ESAC), Villafranca del 
Castillo. P.O. Box  50727. E-28080 Madrid. Spain \email{Pedro.Garcia-Lario@sciops.esa.int}
\and GRAAL, CNRS UMR 5024, Universit\'e de Montpellier 2, F-34095 Montpellier
Cedex 5, France \email{Bertrand.Plez@graal.univ-montp2.fr}  
\and Instituto de Astrof\'{\i}sica de Canarias, La Laguna, E-38200, Tenerife,
Spain \email{amt@iac.es} 
\and Consejo Superior de Investigaciones Cient\'{\i}ficas (CSIC), Spain 
\and Osservatorio Astronomico di Roma, via Frascati 33, I-00040 MontePorzio
Catone, Italy \email{dantona@mporzio.astro.it} 
\and Sterrewacht Leiden, Niels Bohrweg 2, NL-2333 RA Leiden, The Netherlands }

\date{Received 8 June 2006 / accepted 25 August 2006}

\abstract{ Lithium and zirconium abundances (the latter taken as representative
for s-process enrichment) are  determined for a large sample of massive
Galactic O-rich AGB stars,  for which high  resolution optical spectroscopy has
been obtained (R$\sim$40,000$-$50,000). This is done by computing synthetic
spectra based on classical hydrostatic  model atmospheres for cool stars using
extensive line lists.  The results obtained are discussed in  the framework of
``hot bottom burning'' (HBB) and nucleosynthesis models. The complete sample is
studied attending to various observational properties such as the position of the
stars  in the IRAS two-colour diagram ([12] $-$ [25] {\it vs} [25] $-$ [60]),
Galactic distribution, expansion velocity (derived from the OH maser emission)
and period of variability (when  available). We conclude that a considerable
fraction of the sources observed  are actually massive AGB stars ($M>3$--4
$M_{\odot}$)  experiencing HBB, as deduced from the strong Li overabundances
found. A comparison of our results with similar studies carried out in the past
for the Magellanic Clouds (MCs) reveals that, in  contrast to MC AGB stars,
our Galactic sample does  not show any indication of s-process element
enrichment. The differences observed are explained as a consequence of
metallicity effects. Finally, we discuss the  results obtained in the framework
of stellar evolution by  comparing our results with the data available in the
literature for Galactic post-AGB stars and PNe.  
\keywords{Stars: AGB and post-AGB -- Stars: abundances -- Stars: evolution -- Physical data and
processes: Nuclear reactions,  nucleosynthesis, abundances } }

\authorrunning{Garc\'{\i}a-Hern\'andez et al.}
\titlerunning{Li and Zr in massive Galactic O-rich AGBs}
\maketitle
%

\section{Introduction}
The asymptotic giant branch (AGB) is formed by stars with initial masses in the
range between 0.8 and 8 $M_{\odot}$ in a late stage of their evolution. The
internal structure of an AGB star consists of an electron-degenerate C--O core
surrounded by a shell of He. During most of the time H burning is the main
source of energy for the AGB star but, occasionally, the inner He shell ignites
in a ``thermal pulse'' and, eventually, the byproducts of He burning may reach
the outer layers of the atmosphere in a ``dredge-up'' of processed material
(the so-called third dredge-up). Repeated thermal pulses add  carbon to the
stellar surface.  As a consequence, originally O-rich AGB stars  can turn into
C-rich AGB stars (C/O $>$ 1). This scenario  would explain the observed
spectral sequence M--MS--S--SC--C in AGB stars (e.g.\ Mowlavi 1999). 

Another important characteristic of AGB stars is the presence of neutron-rich
elements (s-elements such as Sr, Y, Zr, Ba, La, Nd, Tc, etc.) in their 
atmospheres. These species are formed by slow-neutron captures in the
intershell region. After a dredge-up episode, according to theoretical models,
protons can be partially mixed in the  $^{12}$C- and $^{4}$He-rich region
between the hydrogen- and helium-burning shells (e.g.\ Straniero et al.\ 1995,
1997; Mowlavi 2002; Lattanzio 2003), and can react with $^{12}$C to form
$^{13}$C during the interpulse phase. The $^{13}$C($\alpha$,n)$^{16}$O reaction
releases neutrons, which are captured by iron nuclei and other heavy  elements,
forming s-elements that can later  be dredged up to the stellar surface in the
next thermal pulse (Straniero et al.\ 1995, 1997; Busso et al.\ 2001). $^{22}$Ne
is another neutron source which needs to be considered. In this case the
activation takes place during the convective thermal pulses (e.g.\ Straniero et
al.\ 1995, 2000; Vaglio et al.\ 1999; Gallino et al.\ 2000). $^{22}$Ne can be
formed from $^{14}$N and $^{14}$N is formed by the CNO cycle in the hydrogen
shell. The $^{22}$Ne($\alpha$,n)$^{25}$Mg neutron source requires higher
temperatures and occurs typically at higher neutron densities than the
$^{13}$C($\alpha$,n)$^{16}$O reaction. Thus, a different s-element pattern is
expected depending on the  dominant neutron source. According to the
most recent models the $^{13}$C($\alpha$,n)$^{16}$O reaction is the preferred
neutron source for masses around 1--3 $M_{\odot}$, while for more massive stars
(i.e.\ M $\gtrsim$ 3--4 $M_\odot$)  neutrons are thought to be mainly released
through the $^{22}$Ne($\alpha$,n)$^{25}$Mg reaction  (see, for example, Busso, Gallino
\& Wasserburg 1999 and Lattanzio \& Lugaro 2005 for a recent review). In the
literature, there is  strong evidence that most Galactic AGB stars enriched in
s-process elements have masses around 1--3 $M_{\odot}$, where the
$^{13}$C($\alpha$,n)$^{16}$O reaction is the neutron donor (e.g.\ Lambert et al.\
1995; Abia et al.\ 2001). Unfortunately, a confrontation of the predictions made
by these models with observations of more massive AGB stars in our Galaxy is
not yet available.

In the case of the more massive O-rich AGB stars ($M > 3$--4 $M_\odot$), the
convective envelope can penetrate the H-burning shell activating the so-called
``hot bottom burning'' (hereafter, HBB). HBB takes place when the temperature
at the base of the convective envelope is hot enough ($T \geq
2\times 10 ^{7}$ K) and $^{12}$C can be converted into $^{13}$C and $^{14}$N
through the CN cycle (Sackmann \& Boothroyd 1992; Wood, Bessel \& Fox 1983).
HBB models (Sackmann \& Boothroyd 1992;  D'Antona \& Mazzitelli 1996;
Mazzitelli, D'Antona \& Ventura 1999) also predict the production of  the
short-lived $^{7}$Li by the chain $^{3}$He($\alpha$,$\gamma$)$^{7}$Be
(e$^{-}$,$v$)$^{7}$Li, through the so-called ``$^{7}$Be transport mechanism''
(Cameron \& Fowler 1971). One of the predictions of these models is that
lithium should be detectable, at least for some time, on the stellar surface.

The activation of HBB in massive O-rich AGB stars is supported by  studies
performed of AGB stars in the Magellanic Clouds (hereafter, MCs) (Plez, Smith
\& Lambert 1993; Smith \& Lambert 1989, 1990a; Smith et al.\ 1995) which show a
lack of high luminosity C stars beyond $M_{\rm bol} \sim -6$. Instead, the more
luminous AGB stars in the MCs are O-rich. The detection of strong Li
overabundances together with strong s-element enhancement in these luminous AGB
stars in the LMC and in the SMC is the signature that these stars are indeed
HBB AGB stars that have undergone a series of thermal pulses and dredge-up
episodes in their recent past. The HBB nature of these stars is also confirmed
by the detection of a very small \Crat\ ratio ($\simeq 3$--4), expected only
when HBB is active (Mazzitelli, D'Antona \& Ventura 1999).  Unfortunately, 
current HBB theoretical models have been tested almost  exclusively using the
results obtained from the study of the more massive AGB stars in the MCs but
have never been applied to Galactic sources, mainly because of the lack of
observations  available but also because of the  inaccurate information on 
distances within our Galaxy.

 Indeed, only a handful of Li-rich stars have been found in our Galaxy so far
(e.g.\ Abia et al.\ 1991, 1993; Boffin et al.\ 1993) and, unlike those detected in
the Magellanic Clouds, they are not so luminous ($-6 \leq M_{\rm bol} \leq 
 - 3.5$). Most of them are low mass C-rich AGB stars (Abia \& Isern 1996, 1997,
2000) and intermediate mass S- and SC-stars (Abia \& Wallerstein 1998) and not
O-rich M-type stars. A few are less luminous red giant branch (RGB) stars
at the bump (Charbonnel 2005). Some of these AGB stars are among the most
Li-rich stars in our Galaxy, the so-called ``super Li-rich'' AGB stars (Abia et
al.\ 1991). Under these conditions, however, HBB is not expected to be active
and the Li production is not well understood. Note that HBB is expected to be
active only in the most massive (and luminous) AGB stars (from $\sim$4 to 7
$M_{\odot}$) (Mazzitelli, D'Antona \& Ventura 1999), which should not be
C-rich, but O-rich. Actually, the best candidates in our Galaxy  are the
so-called {\it OH/IR stars},  luminous O-rich AGB stars extremely bright in the
infrared, showing a characteristic double-peaked OH maser emission at 1612 MHz.
These stars are also known to be very long period variables (LPVs), sometimes 
with periods of more than 500 days and large amplitudes of up to 2 bolometric
magnitudes. However, they experience very strong mass loss rates (up to several
times 10$^{-5}$ $M_{\odot}$ yr$^{-1}$) at the end of the AGB and most of them 
appear heavily obscured  by thick circumstellar envelopes, making optical
observations very difficult (e.g. Kastner et al. 1993; Jim\'enez-Esteban et 
al. 2005a; 2006a). Thus, no information yet exists on their lithium abundances
and/or possible s-process element enrichment. 

In this paper we present the results obtained from a wide observational
programme based on high resolution optical spectroscopy of a carefully selected
sample of Galactic AGB stars thought from their observational properties to be massive. 
The criteria followed to select the sources in the
sample are explained in Section 2. The high resolution spectroscopic
observations made in the optical and the data reduction process  are described
in Section 3, while the main results obtained  are presented in Section 4. In
this section we also show how spectral  synthesis techniques were applied to
derive the atmospheric parameters   as well as the lithium and zirconium
abundances of the stars in our sample. A discussion of these results in the
context of HBB models, nucleosynthesis models and stellar evolution can be
found in Section 5. Finally, the conclusions derived from this work are given
in Section 6.


\section{Selection of the sample}
 A large sample of  long period (300--1000 days), large amplitude variability
(up to 8--10 magnitudes in the $V$ band), late-type ($>$ M5) O-rich AGB stars
displaying OH maser emission with a wide range of expansion velocities (from
just a few km s$^{-1}$ to more than 20 km s$^{-1}$) was carefully selected from
the literature. These sources are all very bright  in the infrared  and, as
such, included in the IRAS Point Source  Catalogue (Beichman et al.\ 1988). Some
of them   are so strongly reddened that they do not show any optical
counterpart in the Digitized Sky Survey\footnote{The Digitized Sky Survey was
produced at the Space Telescope Science Institute under US Government grant
NAG W-2166.} (DSS) plates, although in a few cases the detection of  these
sources in the optical at a different epoch has been reported in  the
literature.

\begin{figure*}
\centering
\includegraphics[width=5cm,height=11cm,angle=-90]{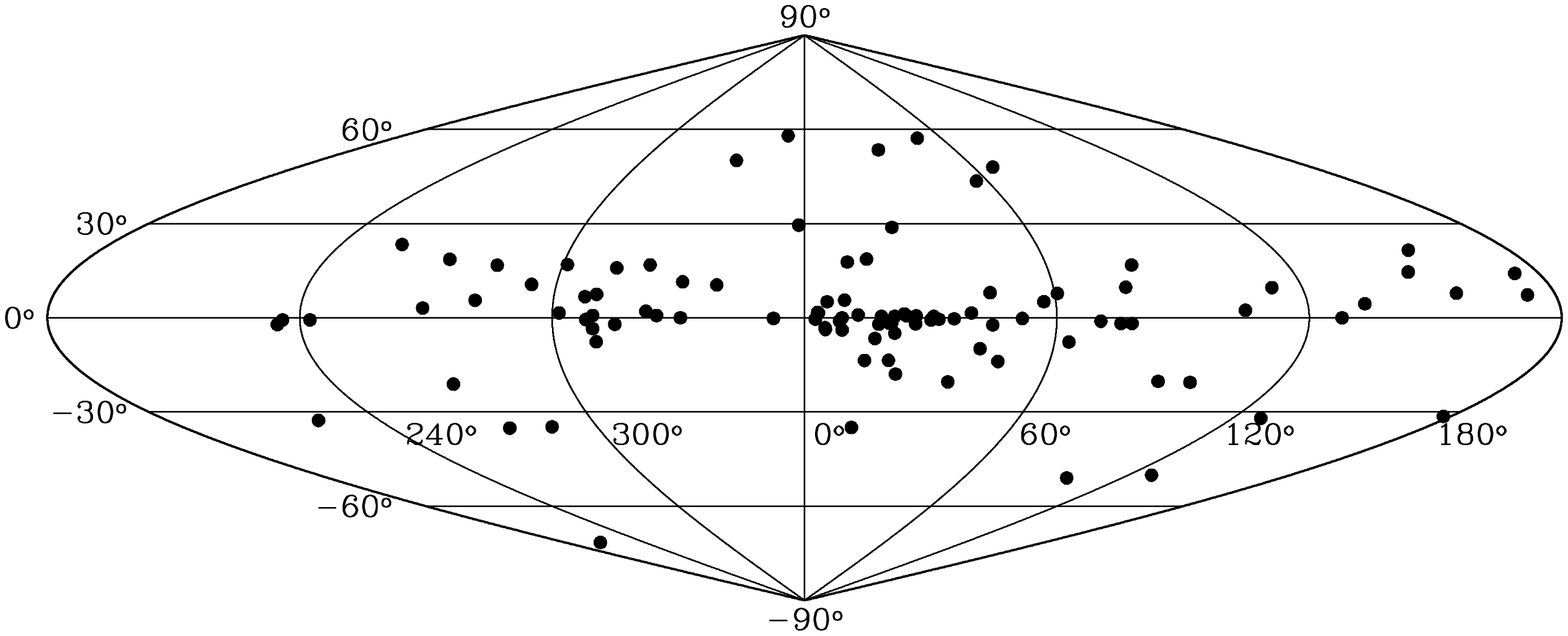}
\caption{Galactic distribution of the sources in the sample.}
\end{figure*}

\begin{figure}
\includegraphics[width=6.5cm,height=8.5cm,angle=0]{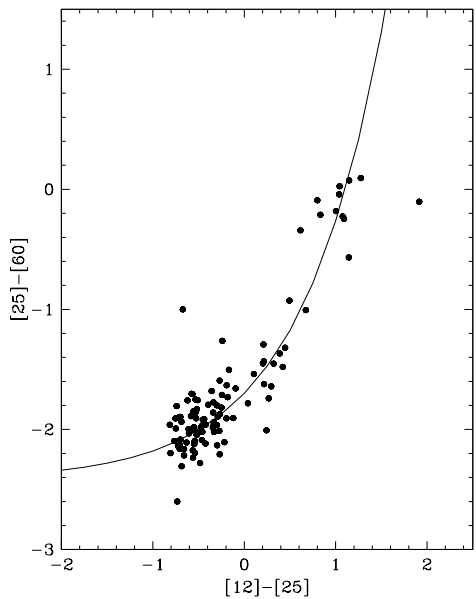}
\caption{IRAS two-colour diagram [12] $-$ [25] {\it{vs}} [25] $-$ [60], where 
the positions of all the sources in the sample are represented (black dots).
The continuous line is the ``O-rich AGB sequence'' (see text) which 
defines the sequence of colours expected for O-rich AGB stars 
surrounded by envelopes
with increasing thickness and/or mass loss rates (Bedijn 1987).}
\end{figure}

Stars were included in the sample if they satisfied at least one of the above
conditions (ideally as many of them as possible), which guarantees that they
are actually relatively massive stars. In Table 1 we list the 102 stars
selected for the analysis, together with their IRAS names, other being names being taken from
the literature, the galactic coordinates, the variability type taken from the
Combined  General Catalogue of Variable Stars (GCVS, Kholopov 1998),  the
spectral type taken from Kwok, Volk \& Bidelman (1997) and references  therein,
and the IRAS colour indices [12] $-$ [25] and [25] $-$ [60]\footnote{Defined as
[12]  $-$ [25] $=$ $-$2.5 log($F_{12}$/$F_{25}$) and [25]$ - $[60] $=$ $-$2.5
log($F_{25}$/$F_{60}$), where $F_{12}$, $F_{25}$ and $F_{60}$ are the IRAS flux
densities at 12, 25 and 60 $\mu$m, respectively.}. Similar information for a
few well known Galactic M-supergiants and S-, SC- and C-type AGB stars that
were also observed for  comparison purposes is presented in Table 2.  In
addition, nine stars classified as Mira-type stars in SIMBAD\footnote{Set of
Identifications, Measurements, and Bibliography for Astronomical Data, operated
at CDS, Strasbourg, France.} and included in the initial sample were found to
exhibit observational  properties corresponding to C-rich stars. These are
listed in Table 3.

The Galactic distribution of the stars in our sample is shown in Figure 1. The
distribution observed clearly suggests that most of them must belong to the
disc population. A few sources with bright optical counterparts  located at
high galactic latitudes may represent a subset of nearby  stars. 

In Figure 2 we show the position of these stars in the  IRAS two-colour diagram
[12] $-$ [25] {\it vs} [25] $-$ [60]. They fall, as expected and with very few
exceptions,  along the observational ``O-rich AGB sequence'' (Garc\'{\i}a-Lario
1992). This sequence agrees very well with the model predictions by Bedijn
(1987) and is interpreted as a sequence of increasing thickness of the
circumstellar envelopes and/or mass loss rates in O-rich AGB stars.

\begin{table*}
\centering
\caption[]{The sample of Galactic O-rich AGB stars}
\footnotesize
\begin{tabular}{llrrcccc}
\hline\hline
IRAS name& Other names &\multicolumn{2}{c}{Galactic Coordinates}&Variability& Spectral 
Type& [12]$-$[25]& [25]$-$[60]\\ 
\hline
01037$+$1219 & WX Psc   &128.64& $-$50.11&Mira & M10   &$-$0.192& $-$1.632 \\
01085$+$3022 & AW Psc   &127.96& $-$32.04&Mira &M9      &$-$0.331  &$-$2.021 \\
01304$+$6211 & V669 Cas &127.81&  $-$0.02&Mira& $\dots$     &$+$0.494& $-$0.927       \\
02095$-$2355 & IRC$-$20029&206.68& $-$71.55&$\dots$ &M6      &$-$0.748  &$-$1.992 \\
02316$+$6455 & V656 Cas &133.61&  $+$4.46&Mira & M8     &$-$0.463& $-$2.089       \\
03507$+$1115 & IK Tau   &177.95& $-$31.41&Mira &M9e     &$-$0.725& $-$2.137  \\
04404$-$7427 & SY Men   &286.97& $-$34.74&Mira &$\dots$      &$-$0.297  &$-$2.132 \\
05027$-$2158 & T Lep    &222.67& $-$32.71&Mira &M7e     &$-$0.706  &$-$2.129 \\
05073$+$5248 & NV Aur   &156.44&  $+$7.83&Mira &M10     &$+$0.204& $-$1.450       \\
05098$-$6422 & U Dor    &274.30& $-$35.16&Mira &M7e     &$-$0.510& $-$1.756  \\
05151$+$6312 & BW Cam   &148.28& $+$14.57&Mira &M9      &$-$0.698& $-$2.084  \\
05559$+$3825 & V373 Aur &173.24&  $+$7.27&Mira &M9$-$10   &$-$0.563  &$-$2.174 \\
06297$+$4045 & IRC$+$40156&174.11& $+$14.12&$\dots$ &M8      &$-$0.094& $-$1.658       \\
06300$+$6058 & AP Lyn   &154.31& $+$21.52&Mira &M7$+$     &$-$0.357  &$-$1.680 \\
07080$-$5948 & NSV 3436 &270.50& $-$21.14&$\dots$ &$\dots$     &$-$0.631& $-$2.109  \\
07222$-$2005 &          &234.58&  $-$2.16&$\dots$ &M9     &$-$0.733  &$-$2.600 \\
07304$-$2032 & Z Pup    &235.90& $-$0.69&Mira&M5e &$-$0.752  &$-$1.908 \\
07445$-$2613 & SS Pup   &242.43&  $-$0.69&Mira &M6e     &$-$0.807& $-$2.197  \\
09194$-$4518 & MQ Vel   &269.09&  $+$3.10&Mira &$\dots$&$-$0.273& $-$2.012       \\
09429$-$2148 & IW Hya   &255.80& $+$23.35&Mira &M9     &$-$0.217  &$-$2.106 \\
10189$-$3432 & V Ant    &271.03& $+$18.61&Mira &M7e     &$-$0.561& $-$2.235  \\
10261$-$5055 & VZ Vel   &281.31&  $+$5.57&SRa &M6e   &$-$0.542& $-$2.096  \\
11081$-$4203 &          &283.71& $+$16.75&$\dots$ &Me      &$-$0.699  &$-$1.895 \\
11525$-$5057 &          &294.00& $+$10.65&$\dots$ &$\dots$      &$-$0.299  &$-$1.964 \\
12377$-$6102 &          &301.63&  $+$1.52&Mira &$\dots$     &$-$0.195  &$-$1.907 \\
12384$-$4536 &          &301.08& $+$16.94&$\dots$ &$\dots$      &$-$0.120& $-$1.905       \\
13203$-$5536 &          &307.45&  $+$6.71&$\dots$ &$\dots$      &$-$0.249& $-$1.818       \\
13328$-$6244 &          &308.00&  $-$0.56&$\dots$&$\dots$   &$+$0.295& $-$1.641       \\
13341$-$6246 &          &308.14&  $-$0.63&L&$\dots$       &$-$0.300& $-$1.799       \\
13379$-$5426 &          &310.14&  $+$7.48&$\dots$ &$\dots$      &$-$0.290& $-$1.893  \\
13442$-$6109 &OH 309.6$+$0.7&309.62&  $+$0.73&$\dots$ &M10     &$-$0.337& $-$1.774  \\
13475$-$4531 & V618 Cen &313.60& $+$15.88&Mira &Me      &$-$0.764& $-$2.096  \\
13517$-$6515 &          &309.51&  $-$3.46&$\dots$ &$\dots$      &$+$0.040& $-$1.782       \\
14086$-$0730 & IO Vir   &334.78& $+$50.12&Mira &M8     &$-$0.333& $-$1.938  \\
14086$-$6907 &          &310.04&  $-$7.61&$\dots$ &$\dots$      &$-$0.333& $-$1.955  \\
14247$+$0454 & RS Vir   &352.67& $+$57.97&Mira &M5e     &$-$0.553& $-$1.847  \\
14266$-$4211 &          &321.67& $+$16.84&$\dots$ &$\dots$      &$-$0.547& $-$2.121  \\
14337$-$6215 &          &314.87&  $-$2.09&$\dots$ &$\dots$      &$-$0.518  &$-$1.831 \\
15099$-$5509 &RAFGL 4212&322.26&  $+$2.10&$\dots$ &$\dots$&$-$0.267 & $-$1.873\\
15193$+$3132 & V S CrB  & 49.47& $+$57.17&Mira &M6e    &$-$0.510 & $-$2.034\\
15211$-$4254 &          &330.43& $+$11.47&$\dots$ &$\dots$      &$-$0.609  &$-$1.997 \\
15255$+$1944 & WX Ser   & 29.51& $+$53.48&Mira &M8e     &$-$0.483  &$-$2.280 \\
15303$-$5456 &          &324.83&  $+$0.69&$\dots$ &$\dots$       &$+$0.421& $-$1.479\\
15576$-$1212 & FS Lib   &358.44& $+$29.52&Mira &M8$-$9    &$-$0.535  &$-$1.860 \\
15586$-$3838 & NSV 7388 &338.79& $+$10.48&$\dots$ &$\dots$      &$-$0.672& $-$1,785  \\
16030$-$5156 & V352 Nor &330.50&  $+$0.01&Mira&M4e     &$-$0.438  &$-$1.914 \\
16037$+$4218 & V1012 Her& 66.88& $+$48.00&$\dots$ &M8      &$-$0.549& $-$1.981  \\
16260$+$3454 & V 697 Her& 56.37& $+$43.53&Mira &M9      &$-$0.245  &$-$1.712 \\
16503$+$0529 & RX Oph   & 23.69& $+$28.78&Mira &M5      &$-$0.687& $-$1.936  \\
17034$-$1024 & V850 Oph & 10.68& $+$17.76&Mira &M8      &$-$0.543& $-$2.007  \\
17103$-$0559 &          & 15.57& $+$18.70 &$\dots$ &$\dots$     &$-$0.539& $-$2.195   \\
17239$-$2812 &          &352.61&  $-$0.18 &$\dots$ &$\dots$     &$-$0.268& $-$1.593   \\
\hline
\end{tabular}
\end{table*}

\begin{table*}
\begin{center}
\footnotesize
\begin{tabular}{llrrcccc}
\hline\hline
IRAS name& Other names &\multicolumn{2}{c}{Galactic Coordinates}&
Variability & Spectral Type & [12]$-$[25]& [25]$-$[60]\\ 
\hline
17359$-$2138 &          &  5.40&  $+$5.09&$\dots$ &M7      &$-$0.705& $-$2.163  \\
17433$-$1750 & GLMP 637 &  9.56&  $+$5.61 &$\dots$&$\dots$      & $+$1.144& $-$0.567   \\
17433$-$2523 &          &  3.10&  $+$1.68 &$\dots$ & $\dots$    &$-$0.421& $-$1.960   \\
17443$-$2519 &RAFGL 5386&  3.29&  $+$1.52 &$\dots$ & $\dots$    & $+$0.322& $-$1.452   \\
17501$-$2656 & V4201 Sgr&  2.58&  $-$0.43 &SR &M8$-$9  &$-$0.239& $-$1.262   \\
18025$-$2113 & IRC$-$20427&  8.93&  $-$0.01&$\dots$ &cM4     &$-$0.167  &$-$1.503 \\
18050$-$2213 & VX Sgr   &  8.34&  $-$1.00&SRc&M4eIa  &$-$0.740& $-$1.805  \\
18057$-$2616 & SAO 186357& 4.89&  $-$3.18&$\dots$ &M       &$-$0.575  &$-$1.703 \\
18071$-$1727 &OH 12.8$+$0.9& 12.75&  $+$0.89 &$\dots$ &$\dots$     &$+$1.275&  $+$0.094   \\
18083$-$2630 & NSV 24316&  4.97&  $-$3.73 &$\dots$ &M6     & $+$0.243& $-$2.007   \\
18172$-$2305 &          &  8.95&  $-$3.89  &$\dots$ & $\dots$ &$-$0.459& $-$2.020   \\
18198$-$1249 &OH 18.3$+$0.4& 18.30&  $+$0.43  &$\dots$ &$\dots$	   & $+$1.092& $-$0.246   \\
18257$-$1000 &V441 Sct  & 21.46&  $+$0.49  &Mira &$\dots$     & $+$1.038& $-$0.042   \\
18273$-$0738 & V436 Sct & 23.73&  $+$1.23  &Mira &$\dots$	   & $+$0.105& $-$1.538   \\
18276$-$1431 & V445 Sct & 17.68&  $-$2.03  &Mira&$\dots$   & $+$1.914& $-$0.103   \\
18304$-$0728 & IRC$-$10434& 24.21&  $+$0.63 &$\dots$ &M5r    &$-$0.394& $-$1.794  \\
18312$-$1209 &          & 20.18&  $-$1.70  &$\dots$ &$\dots$& $+$0.447& $-$1.319   \\
18314$-$1131 & NSV 24500& 20.76&  $-$1.46  &$\dots$ &M6&$-$0.533& $-$1.750   \\
18348$-$0526 & V437 Sct & 26.54&  $+$0.62  &Mira &$\dots$	   & $+$0.615& $-$0.341   \\
18413$+$1354 & V837 Her & 44.56&  $+$8.02 &Mira &M8III   &$-$0.424  &$-$2.118 \\
18429$-$1721 & V3952 Sgr& 16.84& $-$6.60 &Mira &M9	   &$-$0.527  &$-$1.907 \\
18432$-$0149 & V1360Aql & 30.71&  $+$0.42  &Mira &$\dots$      & $+$0.799& $-$0.091   \\
18437$-$0643 & V440 Sct & 26.42&  $-$1.93  &Mira &$\dots$  & $+$0.389& $-$1.367   \\
18454$-$1226 &IRC$-$10463 & 21.53&  $-$4.92 &$\dots$&M6      &$-$0.685& $-$2.306  \\
18460$-$0254 &V1362 Aql & 30.09&  $-$0.68  &Mira &$\dots$       & $+$1.003& $-$0.181   \\
18488$-$0107 &V1363 Aql & 31.99&  $-$0.49  &Mira &$\dots$ 	   &$+$1.043& $+$0.026   \\
18549$+$0208 &OH 35.6$-$0.3& 35.57&  $-$0.34  &$\dots$ &$\dots$	   &$+$0.833& $-$0.212   \\
18560$+$0638 &V1366 Aql & 39.71&  $+$1.50  &Mira &$\dots$       & $+$0.210& $-$1.292   \\
19059$-$2219 & V3880 Sgr& 14.66& $-$13.61  &Mira &M8	   &$-$0.343& $-$1.984   \\
19129$+$2803 &          & 60.67&  $+$7.74 &$\dots$ &$\dots$	   &$-$0.521  &$-$2.045 \\
19147$+$5004 & TZ Cyg  & 81.22& $+$16.80 &Lb&M6    &$-$0.659& $-$2.163  \\
19157$-$1706 &RAFGL 2361& 20.53& $-$13.55 &$\dots$ &$\dots$      &$-$0.458& $-$1.965  \\
19161$+$2343 &RAFGL 2362& 57.12&  $+$5.12  &$\dots$ &$\dots$      & $+$0.217& $-$1.623   \\
19192$+$0922 &NSV 24761 & 44.79&  $-$2.31  &$\dots$ &$\dots$       & $+$0.215& $-$1.433   \\
19254$+$1631 & GLMP 915 & 51.80&  $-$0.22  &$\dots$ & $\dots$	   & $+$1.077& $-$0.226   \\
19361$-$1658 & NSV 24833& 22.74& $-$17.93 &$\dots$ &M8    &$-$0.268  &$-$2.207 \\
19412$+$0337 & V1415 Aql& 42.34&  $-$9.86 &Mira &M9        &$-$0.583& $-$1.889  \\
19426$+$4342 &          & 77.48&  $+$9.73  &$\dots$ & $\dots$	   &$-$0.659& $-$2.216   \\
20052$+$0554 & V1416 Aql& 47.37& $-$13.95 &Mira &M9	   &$-$0.453  &$-$1.919 \\
20077$-$0625 & V1300 Aql& 36.36& $-$20.42  &Mira &M      &$-$0.182& $-$1.731   \\
20109$+$3205 & V557 Cyg & 70.47&  $-$1.08  &Mira &M7	   &$-$0.562& $-$1.876   \\
20181$+$2234 & V371 Vul & 63.44&  $-$7.72  &Mira &$\dots$     & $+$0.268& $-$1.741   \\
20272$+$3535 &GLMP 999  & 75.27&  $-$1.84  &$\dots$ &$\dots$	   & $+$1.147& $+$0.074   \\
20343$-$3020 & RT Mic   & 13.59& $-$34.91 &Mira &$\dots$           &$-$0.605& $-$2.033  \\
20350$+$3741 & V1828 Cyg& 77.90&  $-$1.84 &Mira &M7      &$-$0.814& $-$1.961  \\
22177$+$5936 &NSV 25875 &104.91&  $+$2.41  &$\dots$ &$\dots$	   & $+$0.673& $-$1.006   \\
22180$+$3225 & YZ Peg  &89.60& $-$20.24&$\dots$&$\dots$      &$-$0.620& $-$1.758  \\
22466$+$6942 &          &112.69&  $+$9.59 &$\dots$ &$\dots$          &$-$0.472& $-$1.981  \\
22510$+$3614 & NSV 14347& 97.90& $-$20.56 &$\dots$ &M7        &$-$0.567& $-$1.705  \\
23492$+$0846 & DO 8089& 99.10& $-$51.03&$\dots$&M5          &$-$0.716& $-$1.900  \\
\hline
\end{tabular}
\end{center}
\end{table*}

\begin{table*}
\begin{center}
\caption[]{The sample of comparison stars}
\footnotesize
\begin{tabular}{llrrcccc}
\hline\hline
IRAS name& Other names &\multicolumn{2}{c}{Galactic Coordinates}&Variability& 
Spectral Type & [12]$-$[25]& [25]$-$[60]\\ 
\hline
00213$+$3817& R And  &117.07& $-$23.98    &Mira &S5/4.5e  & $-$0.723 & $-$2.104  \\
02192$+$5821& S Per  &134.62&  $-$2.19    &SRc&M4eIa   & $-$0.407 & $-$1.897  \\
10436$-$3459& Z Ant    &275.86& $+$20.94  &SR &S5.4    & $-$0.830 & $-$2.172  \\
18397$+$1738& V821 Her & 47.78& $+$10.01  &Mira &Ce       & $-$0.873 & $-$1.499  \\
19133$-$1703& T Sgr  & 20.32& $-$13.01    &Mira &S5/6e    & $-$1.125 & $-$1.372  \\
20166$+$3717& WX Cyg   & 75.43&  $+$0.87  &Mira &C9,2eJ   & $-$1.191 &  $+$1.445  \\
23416$+$6130& PZ Cas &115.06&  $-$0.05    &SRc &M3vIa  & $+$0.070 & $-$1.538  \\
23554$+$5612& WY Cas  &115.57&  $-$5.62   &Mira &S6/6e    & $-$0.633 & $-$1.286  \\
23587$+$6004& WZ Cas  &116.76&  $-$1.91   &SRc &SC7/10e& $-$1.280 & $-$1.348  \\
\hline
\end{tabular}
\end{center}
\end{table*}

\begin{table*}
\begin{center}
\caption[]{The sample of peculiar carbon AGB stars}
\footnotesize
\begin{tabular}{llrrcccc}
\hline\hline
IRAS name& Other names &\multicolumn{2}{c}{Galactic Coordinates} &Variability&
 Spectral Type & [12]$-$[25]& [25]$-$[60]\\ 
\hline
00247$+$6922 & V668 Cas &120.86&  $+$6.87  &Mira &C     & $-$0.774 & $-$1.762  \\
04130$+$3918& RAFGL 6312&160.41&  $-$8.09 &$\dots$ &C & $-$1.273 & $-$1.633  \\
09425$-$6040& GLMP 260 &282.04&  $-$5.88  &$\dots$ &C  & $+$0.790 & $-$1.045  \\
17297$+$1747 & V833 Her & 40.80& $+$25.30 &Mira &C   &$-$0.340& $-$1.859   \\
19321$+$2757 & V1965 Cyg& 62.57&  $+$3.96  &Mira &C       & $-$0.703 & $-$1.598  \\
19594$+$4047 & V1968 Cyg& 76.52&  $+$5.59  &Mira &C     & $-$0.285 & $-$1.594  \\
20072$+$3116& V1969 Cyg& 69.35&  $-$0.87  &Mira&C       & $-$0.644 & $-$1.483  \\
23166$+$1655 & LL Peg   & 93.53& $-$40.35  &Mira &C       & $+$0.101 & $-$1.238  \\
23320$+$4316 & LP And   &108.46& $-$17.15  &Mira &C      & $-$0.777 & $-$1.555  \\
\hline
\end{tabular}
\end{center}
\end{table*}

\section{Observations and data reduction}
The observations were carried out during several observing periods in
1996--97. High resolution spectra were obtained using the Utrecht Echelle
Spectrograph (UES) installed at the Nasmyth focus of the 4.2 m William Herschel
Telescope at the Spanish Observatorio del Roque de los Muchachos (La Palma,
Spain) over three different observing runs in August 1996 (run \#1), June 1997
(run \#2) and August 1997 (run \#3) and the CAsegrain Echelle SPECtrograph
(CASPEC) of the ESO 3.6 m telescope at the European Southern Observatory (La
Silla, Chile) in February 1997 (run \#4). The full log of the spectroscopic
observations is shown in Table 4, including more detailed information on the
telescopes, dates of the observations, instrumentation used in each run,
spectral dispersion, as well as on the spectral range covered.

\begin{table*}
\begin{center}
\caption[]{Log of the spectroscopic observations}
\footnotesize
\begin{tabular}{cccccc}
\hline\hline
Run& Telescope &Instrumentation &Dates & Dispersion & Spectral Range\\
& & & & (\AA/pix)& (\AA)\\ 
\hline
\#1 & 4.2 m WHT & UES &16-17 August 1996 &0.065& 5300-9400\\
\#2 & 4.2 m WHT & UES &10-12 June 1997&0.065&4700-10300\\
\#3 & 4.2 m WHT & UES &18-20 August 1997 &0.065&4700-10300\\
\#4 & 3.6 m ESO & CASPEC &21-24 February 1997 &0.085& 6000-8200\\
\hline
\end{tabular}
\end{center}
\end{table*}

We used a TEK 1124 $\times$ 1124 CCD  detector during the first run at the
4.2 m WHT using UES in August 1996 and a SITe1 2148 $\times$ 2148 CCD during the
second and third runs in June 1997 and August 1997, respectively. Since we were
mainly interested in the spectral range between 6000 \AA~and 8200 \AA, we used
the 31.6 line/mm grating in order to provide full coverage of this spectral
region in a single spectrum. With a central wavelength around 6700 \AA~our
spectra extended over $\sim$4000 \AA~during the first run and $\sim$6000
\AA~during the second and third runs. The spectra were taken with a separation
between orders of around 21 pixels (or 7.5$''$), which is large enough to allow sky
subtraction, taking into account that the targets in our sample are
non-extended. The resolving power was around 50\,000, equivalent to a spectral
resolution of 0.13 \AA~around the Li I line at 6708 \AA. The selected set-up
covered the spectral regions from 5300 \AA~to 9400 \AA~in about 47 orders with
small gaps in the redder orders for the first run while the 4700--10300
\AA~region was covered in about 61 orders without any gaps for the second and
third runs.

 CASPEC spectra, taken at the ESO 3.6 m telescope, covered the wavelength range
from 6000 \AA~to 8200 \AA. The red cross-disperser (158 line/mm) was used,
which gave a resolving power of $\sim$40\,000 (equivalent to a spectral
resolution of $\sim$0.17 \AA~around the Li I line at 6708 \AA) and an adequate
interorder separation, using the TEK 1024 $\times$ 1024 CCD. With
this set-up, the selected spectral region is completely covered in  27 orders
with small gaps only between the redder ones. 

The observational strategy was similar in all runs. Since the stars of the
sample are known to be strongly variable, tipically 3--4 magnitudes in the 
$R$-band and sometimes more than 8 magnitudes in the $V$-band, we determined the
exposure time according to the brightness of the source at the telescope. The
typical exposure times ranged from 10 to 30 minutes. Only for a few very faint
stars two 30 minute spectra were taken (and later co-added in order to increase
the S/N ratio of the final spectrum). The goal was to achieve an S/N ratio of
50--150 in the region around 6708 \AA~(the lithium line) in order to resolve
this narrow absorption line, usually veiled at these wavelengths by the
characteristic molecular bands which dominate the optical spectrum of these
extremely cool stars. Note that, because of the very red colours of the sources
observed, the S/N  ratios achieved for a given star can strongly vary from the
blue to the red orders (e.g. 10--20 at $\sim$6000 \AA~while $>$100 at
$\sim$8000 \AA).

At the telescope, all the 102 O-rich stars listed in Table 1 were  tried, but
useful spectra were obtained only for 57 (56\%). The remaining 45 (44\%)
sources were either too red to obtain any useful information on the strength of
the Li I line at 6708 \AA~or the optical counterpart was simply not found, in
the most extreme cases. Some targets with known, relatively bright optical
counterparts in the DSS plates were too faint to be detected  on the date of
the observations. In contrast,  other targets with no optical counterpart
in the DSS plates appeared as considerably bright stars. A few Galactic
M-supergiants together with some C-, SC-, and S-type AGB stars were also
observed for comparison (see list in Table 2). Some of the stars in this latter
group are among the most Li-rich  stars in our Galaxy. In addition, as we have
already mentioned in Section 2, a few stars in the initial sample  turned out
to be carbon stars when observed, and not O-rich as initially suspected (see
list in Table 3). The total number of objects observed was 120.

For wavelength calibration, several Th--Ar lamp exposures were taken every
night during the UES runs. In the case of the CASPEC observations, these
calibration lamp exposures were taken before or after any science exposure at
the position of the target in order to keep control of possible dispersion
changes with the telescope position. Note that with UES the system is more
stable because the instrumentation is located at the Nasmyth focus of the 4.2 m
WHT telescope. Finally, the corresponding bias and flatfield images were also
taken at the beginning (or at the end) of the night. 

The two-dimensional frames containing the echelle spectra were reduced to
single-order one-dimensional spectra using the standard {\sc echelle} software
package as implemented in IRAF.\footnote{Image Reduction and Analysis Facility
(IRAF) software is distributed by the National Optical Astronomy Observatories,
which is operated by the Association of Universities for Research in Astronomy,
Inc., under cooperative agreement with the National Science Foundation.}
Basically, the data reduction process consists of bias-level and 
scattered-light subtraction, the search and trace of apertures using a 
reference bright star, the construction of a normalized flatfield image to
remove pixel-to-pixel sensitivity fluctuations as well as the extraction of the
1-D spectra from the 2-D frames. For the wavelength calibration we selected
non-saturated Th-Ar emission lines and  third or fourth order polynomials for
the fitting.  The calibration accuracy reached was always better than 20
m$\AA$. Finally, we identified the terrestrial features (telluric absorption
lines) comparing the target spectra with the spectrum of a hot, rapidly
rotating star observed on the same night. It should be noted, however,  that
the majority of the spectral ranges used in the abundance analysis
presented in this paper are not significantly affected by these features. 

\section{Results}
\subsection{Overview of the main spectroscopic properties}

\begin{figure*}
\centering
\includegraphics[width=9cm,height=15cm,angle=-90]{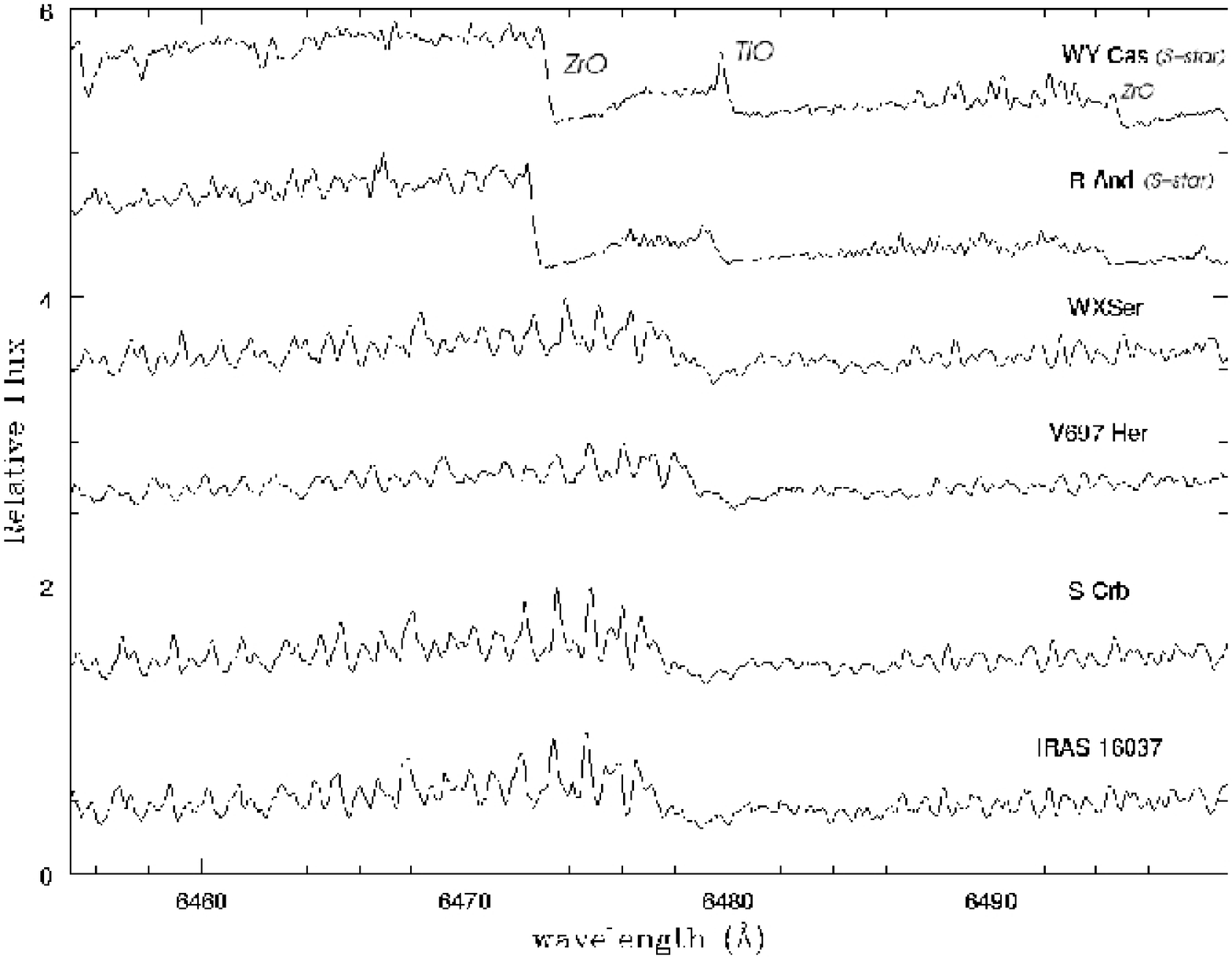}
\caption{High resolution optical spectra of sample stars displaying 
the lack of the ZrO absorption bands at 6474 and 6495 \AA~compared with two
Galactic S-stars (WY Cas and R And), where these bands are prominent. 
WX Ser (IRAS 15255$+$1944) and V697 Her
(IRAS 16260$+$3454) are Li-detected while S CrB (IRAS 15193$+$3132) and IRAS
16037$+$4218 are Li undetected. The absorption band
at $\sim$6480 \AA~corresponds to the TiO molecule.}
\end{figure*}

\begin{figure*}
\centering
\includegraphics[width=9cm,height=15cm,angle=-90]{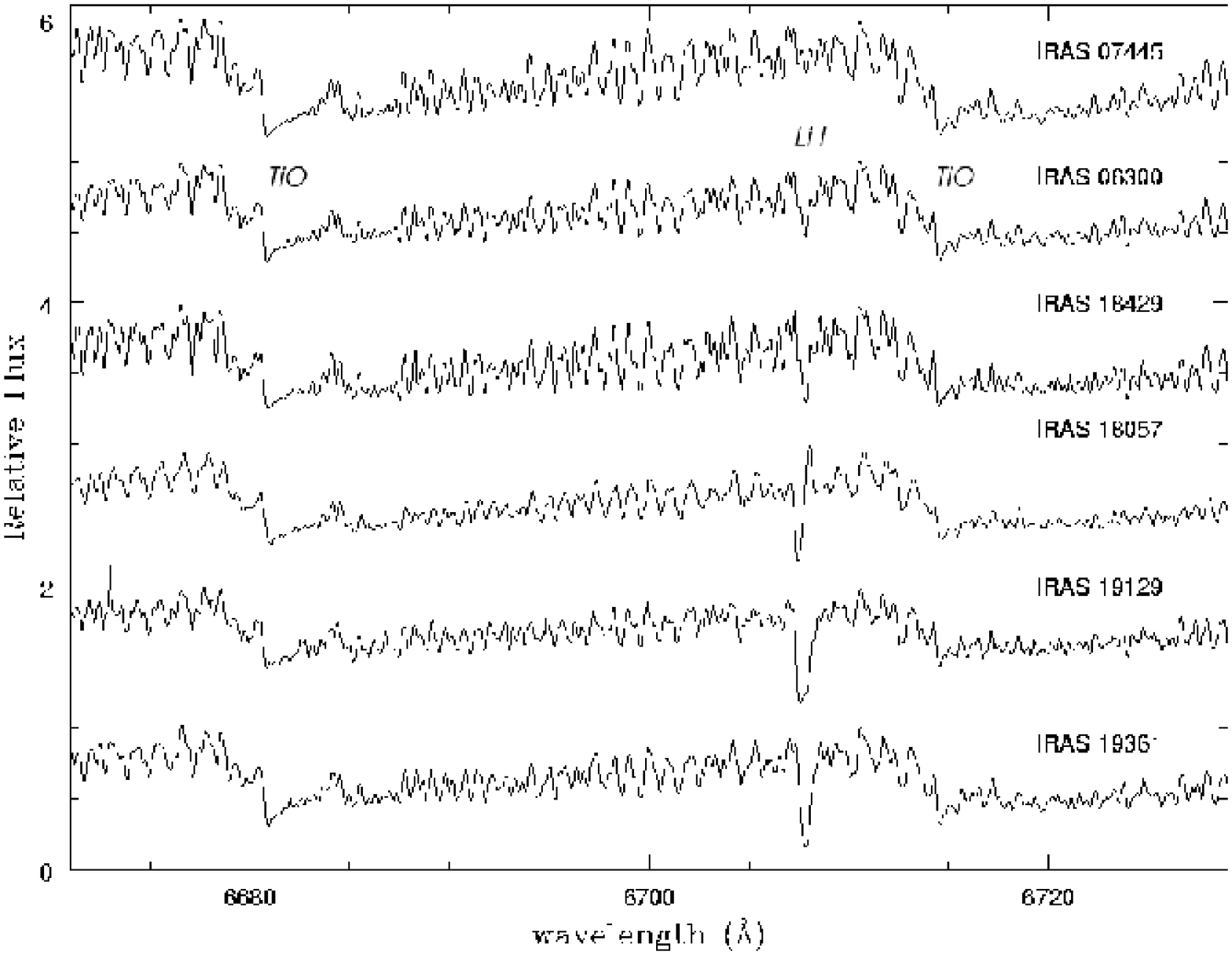}
\caption{High resolution optical spectra of sample stars displaying 
increasing strength of the Li I line at 6708 \AA. Note the P cygni-type
profile of the Li I line in IRAS 18057$-$2616. The jumps at 6681 and 6714
\AA~correspond to bandheads of the TiO molecule.}
\end{figure*}

In general, all stars show extremely red spectra with the flux level falling
dramatically at wavelengths shorter than 6000 \AA. In addition, the spectra are
severely dominated by strong molecular bands mainly due to titanium oxide
(TiO), as a consequence of the very low temperature and the O-rich nature of
these stars.\footnote{In contrast, the spectra of the few C-rich AGB stars found
in our spectroscopic survey (see list in Table 3) are dominated by CN and
C$_{2}$ absorption bands.}  The bandheads of TiO at $\sim$6651, 6681, 6714,
7055, and 7125 \AA~are clearly present in all spectra. Interestingly, the
bandheads of ZrO at $\sim$6378, 6412, 6474, 6495, 6505 and 6541 \AA~seem to be
absent. These ZrO bandheads (together with those corresponding to other
s-element oxides such as LaO or YO) are very strong in Galactic S-stars (see
Figure 3). 
 
 The TiO veiling effect is so intense that it is very difficult to identify
individual atomic lines in the spectra of these stars with the exception of the
Li I line at 6708 \AA, the Ca I lines at 6122 and 6573 \AA, the K I  line at
7699 \AA, the Rb I line at 7800 \AA~and a few strong Fe I lines. The K I and Rb
I resonance lines sometimes show complex profiles,  emission over absorption
and blue-shifted components  which may have their origin in the expanding
circumstellar shell. Actually, the difference between the mean radial
velocities of the stellar and  circumstellar line components is found to be of
the order of the expansion velocity derived from the OH maser emission. In a
few cases the Li I line can also show a similar behaviour (see Figure 4). 
Finally, some stars display H${\alpha}$ emission, which  in this type of stars
is generally interpreted as the consequence of the  propagation of shock waves
through the outer layers of the stellar atmosphere.

\begin{table}[h!]
\begin{center}
\caption[]{Galactic O-rich AGB stars where the lithium line was detected}
\footnotesize
\begin{tabular}{lccccc}
\hline\hline
& & v$_{exp}$(OH)& & Period& \\
IRAS name& Run &(km s$^{-1}$)& Ref.& (days)&Ref. \\ 
\hline
01085$+$3022 & 3  &13.0&3& 560 &1    \\
02095$-$2355 & 1  &$\dots$&  $\dots$ &$\dots$&$\dots$   \\
04404$-$7427$^{*}$ & 4  &7.7 &1&534&2       \\
05027$-$2158 & 4  &7.9&1 & 368 &1    \\
05559$+$3825 & 3  &single peak&4& 590&3\\	    
06300$+$6058 & 3  &12.3&2&440&4      \\
07222$-$2005$^{*}$ & 4  &8.2&1& 1200&10    \\
07304$-$2032 & 4  &7.4&5& 509&1      \\		  
09429$-$2148 & 4  &12.0&2& 650&1     \\
11081$-$4203 & 4  &single peak&1&$\dots$&$\dots$\\
11525$-$5057 & 4  &single peak&1&$\dots$&$\dots$   \\
12377$-$6102$^{*}$ & 4  &20.4&1&$\dots$&$\dots$  \\
14337$-$6215 & 4  &19.7&9&$\dots$&$\dots$  \\
15211$-$4254 & 4  &11.0&1&$\dots$&$\dots$  \\
15255$+$1944 & 1  & 7.4&2&  425&1    \\
15576$-$1212 & 4  &9.9&2&  415 &1    \\
16030$-$5156 & 4  &single peak&1&$\dots$&$\dots$\\
16260$+$3454 & 1  &12.3&2& 475&1     \\
18025$-$2113 & 1  &10.6&1&$\dots$&$\dots$     \\
18057$-$2616 & 1  &18.7&1&$\dots$&$\dots$   \\
18413$+$1354 & 2  &14.7&4 & 590&3 \\
18429$-$1721 & 1  &6.9&1&$\dots$&$\dots$    \\
19129$+$2803 & 2  &single peak&1&420 &10  \\
19361$-$1658 &2,3 &7.5&1&$\dots$ &$\dots$\\
20052$+$0554 & 2  &15.6&12&450 &3\\
\hline
\end{tabular}
\end{center}
$^{*}$ Tentative Li detection\\
\\
 References for the OH expansion velocities: \\
~1) te Lintel Hekkert et al.\ (1991); \\
~2) te Lintel Hekkert et al.\ (1989);\\
~3) Chengalur et al.\ (1993); \\
~4) Slootmaker, Habing \& Herman (1985); \\
~5) Sivagnanam et al.\ (1989); \\
~6) Lewis, Eder \& Terzian (1990);\\
~7) Lewis, David \& Le Squeren (1995);\\
~8) Groenewegen et al.\ (2002) (from CO data);\\
~9) Sevenster et al.\ (1997); \\
~10) Groenewegen \& de Jong (1998); \\
~11) Nyman et al.\ (1992); \\
~12) Lewis (1994)\\
\\
 References for the periods: \\
1) Combined General Catalogue of Variable Stars (GCVS), Kholopov (1998); \\
2) Whitelock et al.\ (1994);\\
3) Jones et al.\ (1990); \\
4) Lockwood (1985);\\
5) Engels et al.\ (1983);\\
6) Nakashima et al.\ (2000); \\
7) Slootmaker, Habing \& Herman (1985);\\
8) Le Bertre (1993); \\
9) Herman \& Habing (1985); \\
10) Jim\'enez-Esteban et al.\ (2006b)\\
\end{table}

\begin{table}[h!]
\begin{center}
\caption[]{Galactic O-rich AGB stars where the lithium line was not detected}
\footnotesize
\begin{tabular}{lccccc}
\hline\hline
& & v$_{exp}$(OH)& & Period& \\
IRAS name& Run &(km s$^{-1}$)& Ref.$^{1}$& (days)&Ref.$^{1}$ \\ 
\hline
03507$+$1115 & 3 &16.5&2&470 &1\\
05098$-$6422 & 4 & 6.0&2&394 &2\\
05151$+$6312 & 1 &14.5&2&$\dots$&3\\
07080$-$5948 & 4 &5.0&1&$\dots$&$\dots$\\
07445$-$2613 & 4 &4.8&2& 391 &1\\
10189$-$3432 & 4 &3.4&2& 303 &1\\
10261$-$5055 & 4 &3.7&1&317&1\\
13379$-$5426 & 4 &10.2&1&$\dots$&$\dots$\\
13442$-$6109$^{**}$ & 4 &23.7&1&$\dots$&$\dots$\\	     
13475$-$4531 & 4 &3.7&1&$\dots$&$\dots$\\
14086$-$0730$^{*}$ & 4 &13.7&2& $\dots$&$\dots$\\
14086$-$6907$^{*}$ & 4 &13.4&1&$\dots$&$\dots$\\
14247$+$0454 & 4 &4.0&2& 354 &1\\
14266$-$4211 & 4 &9.1&1&$\dots$&$\dots$\\
15193$+$3132 & 1 &3.3&2& 360 &1\\
15586$-$3838 & 4 &9.3&1&$\dots$&$\dots$\\
16037$+$4218 & 1 &3.5&1& 360 &10\\
16503$+$0529 & 2 &3.8&3&323 &1\\
17034$-$1024 & 2 &single peak&1&346 &1\\
17359$-$2138$^{*}$ & 2 &8.5&1&$\dots$&$\dots$\\
18050$-$2213 & 3 &19.8&2&732 &1 \\
18304$-$0728 & 3 &15.3&2&$\dots$&$\dots$\\
18454$-$1226$^{**}$ & 3 &single peak&1&$\dots$&$\dots$\\
19147$+$5004 & 1 &12.6&1&$\dots$&$\dots$\\
19157$-$1706 & 2 &14.7&1&$\dots$&$\dots$\\
19412$+$0337 & 2 &14.5&6& 440 &3\\
20343$-$3020 & 3 &8.4&1&349&2\\
20350$+$3741 & 3 &$\dots$&$\dots$ & 450 &3\\
22180$+$3225 & 1 &4.3&3&$\dots$&$\dots$\\
22466$+$6942 & 1 &5.6&1&408&6\\
22510$+$3614 & 1 &2.8&3&$\dots$&$\dots$\\
23492$+$0846$^{**}$ & 1 &$\dots$&$\dots$&$\dots$&$\dots$\\
\hline
\end{tabular}
\end{center}
$^{*}$ Poor S/N ratio\\
$^{**}$ Possibly not an AGB star\\ 
\\
$^{1}$ See applicable references in Table 5\\
\end{table}

\begin{table}[h!]
\begin{center}
\caption[]{Galactic O-rich AGB stars for which no analysis of the lithium line was possible}
\footnotesize
\begin{tabular}{lcccccc}
\hline\hline
& & v$_{exp}$(OH)& & Period& &\\
IRAS name& Run &(km s$^{-1}$)&Ref.$^{1}$& (days)&Ref.$^{1}$ & Notes\\ 
\hline
01037$+$1219 & 1,3 &18.2&2&660 &2& R\\
01304$+$6211 & 3 &11.0&2& 1994 &7& $\dots$\\
02316$+$6455 & 3 &15.5&7&534 &3& R\\
05073$+$5248 & 3 &16.9&2& 635 &1& R\\
06297$+$4045 & 3 &12.0&2&  520 &10& R\\
09194$-$4518 & 4 &11.3&1&$\dots$ &$\dots$& R\\ 
12384$-$4536 & 4 &13.2 &1&$\dots$&$\dots$& R\\
13203$-$5536 & 4 &12.7&1 &$\dots$&$\dots$& $\dots$\\ 			      
13328$-$6244 & 4 &25.1&1&$\dots$&$\dots$& $\dots$\\
13341$-$6246 & 4 &16.7&1&$\dots$&$\dots$& R\\ 
13517$-$6515 & 4 &17.9&1&$\dots$&$\dots$& $\dots$\\
15099$-$5509 & 4 &18.8&1&$\dots$&$\dots$& R\\
15303$-$5456 & 4 &24.0&1&$\dots$&$\dots$& $\dots$\\	
17103$-$0559 & 2,3 &15.3&1&  420 &10& $\dots$\\
17239$-$2812 & 3 &20.7&1&$\dots$&$\dots$& $\dots$\\
17433$-$1750 & 3 &15.0&1&$\dots$ &$\dots$ & R\\
17433$-$2523 & 3 &19.9&1&$\dots$&$\dots$& R\\
17443$-$2519 & 3 &16.1&1&$\dots$&$\dots$& $\dots$\\
17501$-$2656 & 3 &23.0&1&928 &8& R\\
18071$-$1727 & 3 &11.0&2& 1488 &9& $\dots$\\
18083$-$2630 & 2,3 &18.2&2&$\dots$&$\dots$& $\dots$\\	 
18172$-$2305 & 3 &15.9&1&$\dots$&$\dots$& $\dots$\\
18198$-$1249 & 3 &15.3&2&  845 &9& $\dots$\\
18257$-$1000 & 2,3 &19.1&2& 1975 &9& $\dots$\\
18273$-$0738 & 2 &14.3&2& 700 &5& R\\
18276$-$1431 & 2 &12.0&2&  890 &9& R\\
18312$-$1209 & 3 &12.3&1&$\dots$&$\dots$& $\dots$\\
18314$-$1131$^{*}$ & 3 &$\dots$ &1&$\dots$&$\dots$& R\\
18348$-$0526 & 2,3&13.9&2& 1575 &3& $\dots$\\
18432$-$0149 & 2,3 &17.5&2& 1140 &5& $\dots$\\
18437$-$0643 & 2,3 &12.1&2&  652 &9& $\dots$\\
18460$-$0254 & 2,3 &20.2&2& 1730 &5& $\dots$\\
18488$-$0107 & 3 &20.5&1&1540&4& $\dots$\\
18549$+$0208 & 3 &13.7&1&  840 &9& $\dots$\\
18560$+$0638 & 1,2,3 &16.0&2& 1033 &5& $\dots$\\
19059$-$2219 &2,3&13.3&2&  510 &1& R\\
19161$+$2343 &2,3 &17.9&2&$\dots$&$\dots$& $\dots$\\
19192$+$0922 & 2,3 &16.5&2&  552 &9& $\dots$\\
19254$+$1631 & 2 &19.0&2& 1162 &9& R\\ 
19426$+$4342 &1,3&8.6&1&$\dots$&$\dots$& R\\ 
20077$-$0625 & 1,3 &12.3&2&  680 &1& R\\
20109$+$3205 & 3 &5.7&1&  382 &1& $\dots$\\
20181$+$2234 & 1,3 &11.2&2&$\dots$&$\dots$& $\dots$\\
20272$+$3535 & 3 &12.0&2& 1603 &9& $\dots$\\
22177$+$5936 & 3&15.1&2& 1215 &9& $\dots$\\ 
\hline
\end{tabular}
\end{center}
$^{*}$ Single peak OH maser\\
R: optical counterpart too red; \\
$\dots$: optical counterpart not found\\
\\
$^{1}$ See applicable references in Table 5\\
\end{table}

\begin{table}[h!]
\begin{center}   
\caption[]{Comparison stars}
\footnotesize
\begin{tabular}{lccccc}
\hline\hline
& & v$_{exp}$(OH)& & Period& \\
IRAS name& Run &(km s$^{-1}$)& Ref.$^{1}$& (days)&Ref.$^{1}$ \\ 
\hline
00213$+$3817& 1 &9.0&10& 409 &1\\
02192$+$5821& 1 &14.5&2& 822 &1\\
10436$-$3459& 4 &11.3&1&104&1\\
18397$+$1738& 2 &13.4 &8 & 511 &3\\
19133$-$1703& 1 &10&10& 395 &1\\
20166$+$3717& 3 &$\dots$&$\dots$ & 410 &1\\
23416$+$6130& 1 &26.5&2& 925&1\\
23554$+$5612& 1 &15.5&10& 477 &1\\
23587$+$6004& 1 &$\dots$&$\dots$&186 &1\\
\hline
\end{tabular}
\end{center}
$^{1}$ See applicable references in Table 5\\
\end{table} 

\begin{table}[h!]
\begin{center}
\caption[]{Peculiar C-rich AGB stars}
\footnotesize
\begin{tabular}{lcccccc}
\hline\hline
& & v$_{exp}$(CO)& & Period& &\\
IRAS name& Run &(km s$^{-1}$)&Ref.$^{1}$& (days)&Ref.$^{1}$ & Notes\\ 
\hline
00247$+$6922 & 3 &17.2&8 &650&3& $\dots$\\
04130$+$3918& 3 &$\dots$&$\dots$& 470  &10 & B\\
09425$-$6040& 4 &$\dots$&$\dots$&$\dots$&$\dots$ & B\\
17297$+$1747 & 2 &15.2&11& 520&3&R\\
19321$+$2757 & 2 &24.4 &8&625 &3& R \\
19594$+$4047 & 3 &20.5 &8 &  783 &3& $\dots$\\
20072$+$3116& 2 &25.6  &8& 550 &3& R\\
23166$+$1655 & 3 &15.1 &8&  520 &3& $\dots$\\
23320$+$4316 & 3 &14.7 &8&  620 &3& R\\
\hline
\end{tabular}
\end{center}
B: bright optical counterpart; \\
R: optical counterpart too red; \\
$\dots$: optical counterpart not found\\
\\
$^{1}$ See applicable references in Table 5\\
\end{table} 

We detected the presence of the Li I resonance line at 6708 \AA~in 25\% of the
sources in the sample (25 stars) with a wide variety of strengths, while  in
31\% of them (32 stars) we did not find any signature of this line. The
remaining 44\% (45 stars) were either too red or the optical counterpart was simply 
not found at the moment of the observations. Sample spectra of stars showing
increasing strength of the Li I line at 6708 \AA~ are presented in Figure 4.

The observed sample has been divided into two  different groups on the basis of
the detection or non-detection of the lithium line in Tables 5 and 6,
respectively. A third group is formed by the stars that were too red (or which
did not show any optical counterpart at the telescope), and these are listed in
Table 7. The IRAS name, the run in which the stellar field was observed, the OH
expansion velocity and the pulsational period (if available) are given for
every source in these tables. The same information is also presented for the
comparison stars in Table 8 and  for the sample of peculiar C-rich AGB stars in
Table 9 (in this latter  case the expansion velocities are derived from CO
data).

The detection of strong circumstellar Li I and Rb I lines in S Per (IRAS
02192$+$5821), sometimes classified in the literature as an M-type Galactic
supergiant, is remarkable. M-type supergiant stars are not expected to show a
strong Li I line (e.g. Luck \& Lambert 1982) or any s-process element
enhancement (e.g. Smith et al. 1995). According to our data, S Per looks like a
possible genuine O-rich AGB star. In addition, we find a cool effective
temperature of $\sim$3000 K which is quite different from Levesque et al. (2005)
and further investigation is needed. Similarly, PZ Cas (IRAS 23416$+$6130),
another source usually classified as an M-type supergiant (e.g. Levesque et al.\
2005), is for the first time clearly identified by us as a possible S-,SC-type
AGB star. PZ Cas seems to be slightly enriched in Zr and displays some strong Ba
atomic lines but it is not enriched in Li, as expected if PZ Cas is a low mass
S-,SC-type AGB star. Consistent with our possible AGB identification for the
latter two sources, they show variability amplitudes of more than 3 magnitudes
in the $V$-band and are the only two stars with periods well beyond 500 days in
Table 8. A more detailed analysis of these two stars is beyond the scope of this
paper but it would be needed to reach a definitive conclusion. Also interesting
is the new detection of a strong lithium line in the peculiar mixed chemistry
(C-rich and O-rich) star IRAS 09425$-$6040, which is studied in  a separate
paper (Garc\'\i a-Hern\'andez et al.\ 2006a).  Lithium was also detected in the
C-rich AGB stars IRAS 04130$+$3918, IRAS 20072$+$3116 (V1969 Cyg) and IRAS
23320$+$4316 (LP And), for which  a detailed analysis will be presented
elsewhere.  Finally, the S-star IRAS 10436$-$3459 (Z Ant), observed for the
first time with high resolution spectroscopy, shows, as expected, intense
molecular bands of s-element oxides  (like ZrO and LaO) but no lithium. This
star will also be analysed in  detail in a separate publication. 

\subsection{Modelling strategy}
Atmospheres of cool pulsating AGB stars present a major challenge for
realistic, self-consistent modelling. At present, two types of self-consistent
models exist. Classical hydrostatic model atmospheres include a sophisticated
treatment of micro-physical processes and radiative transfer but neglect the
effects of dynamics due to pulsation and winds (e.g.\ Plez, Brett \& Nordlund
1992). Time-dependent dynamical models for stellar winds (Winters et al.\ 2000;
H\"{o}fner et al.\ 2003, and references therein) include pulsation, dust
formation but a more crude description of radiative transfer. Propagating shock
waves caused by stellar pulsation modify the structure of the atmosphere on
local and global scales, which give rise to strong deviations from a
hydrostatic stratification. The radiative field is dominated by the effect of
molecular opacities or even by the dust grains forming in the cool outer layers
of these atmospheres. Chemistry and dust formation may be severely out of
equilibrium. 

Hydrostatic model atmospheres based on the MARCS code (Plez, Brett \& Nordlund
1992; Gustafsson et al.\ 2003) reproduce well  the observed spectra of C-rich
AGB stars (e.g.\ Loidl, Lancon \& J{\oe}rgensen 2001) and O-rich AGB stars (e.g.\
Alvarez et al.\ 2000) in the optical domain and were thus adopted for the
analysis. Indeed, the optical spectra of massive O-rich AGB stars in the
Magellanic Clouds have  succesfully been modelled using this code (e.g.\ Plez,
Smith \& Lambert 1993). A comparison of these  models with high resolution
observational data in the optical for massive O-rich AGB stars in our Galaxy 
is unfortunately still lacking.

Our analysis combines state-of-the-art line-blanketed model atmospheres and
synthetic spectroscopy with extensive line lists. For this we need to  estimate
first the  range of values of the stellar parameters that can reasonably be
adopted for the stars in our sample: effective temperature {$T_{\rm eff}$},
surface gravity {log $g$}, mass {$M$}, metallicity
{$Z =$ [Fe/H]}, microturbulent velocity {$\xi$} and {C/O}
ratio. 

Spherically symmetric, LTE, hydrostatic model atmospheres were calculated using
the MARCS code for cool stars (Gustafsson et al.\ 2003).  The models are
designed with the notation {($T_{\rm eff}$, log $g$, $M$, $Z$, $\xi$, C/O)}, where
$T_{\rm eff}$ is the effective temperature of the star in K, \textit{g} is
the surface gravity in cm s$^{-2}$, \textit{M} is the stellar mass in
$M_\odot$, \textit{Z} is the metallicity, \textit{$\xi$} is the microturbulent
velocity in km s$^{-1}$ and {C/O} is the ratio between the C and O
abundances. 

 Synthetic spectra were generated with the TURBOSPECTRUM package  (Alvarez \&
Plez 1998) which shares much of its input data and routines with MARCS. Solar
abundances of Grevesse \& Sauval (1998) were always adopted, except for the
iron  abundance, which was taken as {log $\varepsilon$(Fe)}$ =$ 7.50. 
Most opacity sources for cool M-type stars were included and taken from the
literature. In addition,  updated line lists were produced by us for TiO (Plez
1998), ZrO (Plez et al.\ 2003) and VO (Alvarez \& Plez 1998).

For atomic lines, the primary source of information was the VALD-2 database
(Kupka et al.\ 1999). The NIST\footnote{National Institute of Standards and
Technology (NIST);   http://physics.nist.gov/cgi-bin/AtData/lines$\_$form.}
atomic spectra database was also consulted for comparison. Where  possible,
\textit{gf} values were checked by fitting the solar spectrum. VALD-2
\textit{gf} values taken from Kurucz (1993, 1994) were often erroneous and
did not fit the atomic lines of the solar spectrum.  The identification of
features was made using the solar atlases by Moore, Minnaert, \& Hootgast
(1966) and Wallace, Hinkle \& Livingston (1993, 1998) and the observed solar
spectrum of Neckel (1999). The TURBOSPECTRUM program was run using the
abundances from Grevesse \& Sauval (1998) together with the solar model
atmosphere by Holweger \& M\"{u}ller (1974) with parameters
{$T_{\rm eff}$} = 5780 K, {log $g$} = 4.44 and a variable microturbulence
as a function of the optical depth. 

The whole machinery was tested on the high resolution optical spectrum
($R\sim$ 150\,000) of the K2 IIIp giant Arcturus ($\alpha$ Boo) from Hinkle et
al.\ (2000). A MARCS model atmosphere was used with the fundamental parameters
determined by Decin et al.\ (2003a): {$T_{\rm eff}$} = 4300 K, {log
$g$} = 1.50, \textit{Z} = $-$0.5, \textit{M} = 0.75  $M_\odot$, {$\xi$} = 1.7 km
s$^{-1}$, {C/N/O} = 7.96/7.55/8.67 and {$^{12}$C/$^{13}$C} = 7. 
These parameters are in excellent agreement with other determinations reported
in the literature. Some Kurucz \textit{gf} values of metallic  lines of Fe I,
Ni I, Co I, Cr I, Ti I, Zr I and Nd II, which were too weak in the solar
spectrum, were adjusted so as to yield a good fit to the Arcturus spectrum in
several regions (especially in the 7400--7600 \AA~and 8100--8150 \AA~regions);
otherwise we used the \textit{gf} values from the VALD-2 database. This exercise
was very useful in confirming the lack of s-process atomic lines in the
7400--7600 and 8100--8150 \AA~spectral windows in the Galactic O-rich AGB
stars in our sample, as we shall see later.

\subsection{Spectral regions of interest}
For the abundance analysis we concentrated our attention on the following
spectral regions:

\subsubsection{Lithium}  
We used the Li I resonance line at $\sim$6708 \AA~to derive the lithium
abundances, which can be used as a signature of HBB.  The selected spectral
region ranges from 6670 to 6730 \AA~ and covers the TiO molecular bandheads at
$\sim$6681 and 6714 \AA, which are sensitive to variations in the effective
temperature.  In addition, the spectral regions around the subordinate Li I
lines at $\sim$6103 and 8126 \AA~were also inspected. However, we decided to
drop  them from the analysis because the latter lines are much weaker.
Moreover, the S/N ratio is usually very low around 6103 \AA~in our sources and
the contamination by telluric lines at 8126 \AA~is very strong. 

\subsubsection{Rubidium and Potassium} 
We also synthesized the spectral regions around the resonance lines of Rb I at
$\sim$7800 \AA~and K I at 7699 \AA, with the intention of using the elemental
abundances derived from these lines as neutron density and  metallicity
indicators, respectively. In addition, these spectral regions were also  used
to check that the model parameters adopted to reproduce the lithium  region
also provided a reasonably good fit at other wavelengths.  Two  intervals
covering $\sim$60 \AA~in the regions 7775$-$7835 \AA~and  7670$-$7730 \AA~were 
selected for this purpose. Unfortunately, the frequent detection of 
circumstellar components to these lines prevented the accurate determination 
of the K I elemental abundances, and they will not be discussed here. To a
lesser extent the problem also affects the Rb I line. A  detailed analysis of
the Rb abundances  will be presented in a forthcoming paper (Garc\'\i
a-Hern\'andez  et al.\ 2006b). 

\subsubsection{s-process elements} 
The study of the s-process elements was carried out through the analysis of
several ZrO absorption bands in the region from 6455 to 6499 \AA. In
particular, we used the ZrO bandhead around  $\sim$6474 \AA\ to determine 
severe upper limits to the Zr elemental abundance, taken as  representative of
the overall s-process element enrichment.  Surprisingly, this band was not
detected in any star in our sample, neither did we find any signature from atomic
Zr or from atomic lines corresponding to any other abundant s-process element 
in the  7400$-$7600 \AA~and 8100$-$8150 \AA\ spectral windows, where many atomic
lines of Zr I, Nd II, Ba II might potentially have been found, as we show
later.

\subsection{Derivation of the stellar parameters}
In order to analyze how the variations in stellar parameters influence the
output synthetic spectra, we constructed a grid of MARCS model atmospheres and
generated the associated synthetic spectra with the following specifications:
i) the stellar mass was in all cases taken to be 2 $M_\odot$;  ii)
{$T_{\rm eff}$} values ranging from 2500 to 3800 K in steps of 100 K; iii)
{C/O} ratio values of 0.5, 0.7, 0.8 and 0.9; iv) {log $g$} between
$-$0.5 and 1.6 dex in steps of 0.3 dex; v) microturbulent velocity
{$\xi$} between 1 and 6 km s$^{-1}$ in steps of 0.5  km s$^{-1}$; vi)
metallicity \textit{Z}  between 0.0 and $-$0.3 dex; vii) {log
$\varepsilon$(Zr)}\footnote{The zirconium abundance is given in the usual scale
12 $+$ {log $N$(Zr)}. On this scale the solar zirconium abundance is 2.6
dex.} from 1.6 to 3.6 dex in steps of 0.25 dex; viii) CNO abundances shifted
$\pm$1.0 dex (in steps of 0.5 dex) from the solar values of Grevese \& Sauval
(1998); and ix) {$^{12}$C/$^{13}$C} ratios of 10 (as expected for HBB
stars) and 90 (the solar value). Finally, the synthetic spectra were convolved
with a Gaussian profile (with a certain {FWHM} typically between 200 and
600 m\AA) to account for macroturbulence as well as instrumental profile
effects.

This, however, results in an extremely large grid of synthetic spectra
containing thousands of possible combinations of the above
parameters! In order to reduce the number of spectra to be considered in our
analysis, further constraints were imposed by studying the sensitivity of
our spectral synthesis to changes in these parameters.

\subsubsection{Stellar mass}
As we have already mentioned, the stellar mass was in all cases 
considered to be constant and equal to 2 $M_\odot$. 
This is because the temperature and
pressure structure of the model atmosphere is practically identical for a 1
$M_\odot$ and a 10 $M_\odot$ model atmosphere (see figure 1 in Plez 1990). 
Increasing the mass will just have a marginal effect on the 
output synthetic spectrum through a decrease, for a given gravity, of the
atmospheric extension.

\subsubsection{Effective temperature} 
The synthetic spectra are, in contrast, particularly sensitive to 
{$T_{\rm eff}$}, which determines the overall strength of the molecular
absorption over the continuum (mainly, from TiO molecules but also from VO, and
from ZrO if the Zr elemental abundance is increased above a certain limit).
Thus, we decided to keep the whole range of values initially considered in our
grid, i.e.\ {$T_{\rm eff}$} from 2500 to 3800 K in steps of 100 K.
Actually, the depth of the molecular bands increases considerably with
decreasing temperature. The effective temperature has also a large impact on
the strength of the Li I line at 6708 \AA~in the  synthetic spectrum. A
decrease of {$T_{\rm eff}$} has the effect of  increasing the TiO veiling,
as well as the strength of the Li I line, as a consequence of the changes in
the Li I/Li II population equilibrium.

\subsubsection{C/O ratio}
The C/O ratio basically determines the prevalence of O-rich molecules
{C/O} $<$ 1) against C-rich molecules like CN, C$_{2}$, etc. In the case
of our O-rich AGB stars, the precise value of the  {C/O} ratio adopted,
always $<$ 1, affects mainly the strength of the TiO veiling and of the Li I
line in the synthetic spectrum. An increase of the  {C/O} ratio will
decrease the TiO veiling and increase the Li I line strength relative to the
adjacent continuum in a very similar way as a decrease in  {$T_{\rm eff}$}.
This is a good example of how a set of parameters providing a good match to the
observations  for a given spectral region is not necessarily unique. A good fit
may be obtained for another combination of parameters that can correspond to
quite different Li abundances. Fortunately, one of these two sets of
parameters  usually does not provide acceptable fits for other spectral
regions. In particular, the presence of detectable atomic lines in the
synthetic spectrum is very sensitive to variations in the {C/O} ratio:
an increase in the {C/O} ratio will lower the TiO veiling, making the
detection of atomic lines easier. According to the models, the immediate effect should
be the detection of much stronger atomic lines of  K I, Fe I, Zr I, Nd II, etc.
Since we do not see any such effect  in our spectra, we conclude that {C/O}
must always be $\leq$ 0.75 in our stars. Abundances derived from models with
{C/O} between 0.15 and 0.75 show actually little differences (Plez,
Smith \& Lambert 1993). Given that all stars in our sample are clearly O-rich
and taking into account the above considerations, we decided to use
{C/O} = 0.5 in all cases, as a constant value. This selection is in
agreement with other more detailed determinations made in massive O-rich AGB
stars in the MCs (e.g.\ Plez, Smith \& Lambert 1993; Smith et al.\ 1995) and in a
few low mass M-type stars studied in our Galaxy (Smith \& Lambert 1985,
1990b).\footnote{This is the only other chemical analysis previously carried out
in Galactic O-rich AGB M-type stars (only six stars!) before the work presented
here. However, these were low mass stars ($M \lesssim  1.5$--$2 M_\odot$) and
their study was focussed on the determination of CNO abundances through near-IR
spectra and the s-process enrichment, but they did not study the lithium
enrichment in particular.} 

\subsubsection{Surface gravity}
The surface gravity also affects the appearance of the output synthetic spectra
but its effect is small compared to that of the effective temperature. The
influence of a change in the value adopted for the surface gravity within the
range of values ({log $g$} between $-$0.5 and 0.5 dex) under consideration
is not as severe as a temperature change. The molecular absorption becomes
slightly weaker at higher surface gravities. For example, an increase in the
surface gravity of 0.5 dex has approximately the same impact as an increase of
100 K in effective temperature. However, for a fixed temperature we found that
the lowest gravities in general fit better both the TiO band strength and the
pseudo-continuum around the 6708 \AA~Li I line. As the appearance of the spectra
is not so dependent on the surface gravity, and considering that its value must
be low in these mass-losing stars, a constant surface gravity of {log
$g$} = $-$0.5 was adopted for all the stars in the sample. This selection seems to
be also appropiate if it is compared with the values adopted by Plez, Smith \&
Lambert (1993) of $-$0.33 $\leq$ {log $g$} $\leq$ 0.0 for O-rich AGB stars
in the MCs with effective temperatures between 3300 and 3650 K. 

\subsubsection{Microturbulent velocity}
The microturbulent velocity is usually derived by demanding no correlation
between the abundance of iron derived from individual lines and their reduced
equivalent widths. Previous studies of O-rich AGB stars found microturbulent
velocities between 2 and 4 km s$^{-1}$ (Plez, Smith \& Lambert 1993; Smith \&
Lambert 1985, 1989; Smith et al.\ 1995; Vanture \& Wallerstein 2002).
Unfortunately, we cannot estimate the microturbulent velocity in the stars of
our sample due to the lack of useful Fe I atomic lines (the small number of
detectable Fe I lines does not cover a wide range of equivalent widths). Thus,
we will assume in the following a microturbulent velocity $\xi$ = 3 km s$^{-1}$,
which is a typical value generally adopted for AGB stars (e.g.\ Aringer,
Kerschbaum \& J$\o$rgensen 2002). Higher values of $\xi$ would strenghten all
atomic lines in the synthetic spectra and we do not see this effect in our
data. 

\subsubsection{Metallicity}
The metallicity of the Galactic O-rich AGB stars in our sample is  assumed to
be solar, as expected for stars belonging to the disc population of our Galaxy.
A lower metallicity is unlikely since all the stars are expected to be at least
of intermediate mass ($M \gtrsim  2$--$3\ M_\odot$), if not more massive.  This
assumption is also in good agreement with the typical metallicities  derived
for other AGB stars in our Galaxy (e.g.\ Abia \& Wallerstein 1998; Vanture \&
Wallerstein 2002). The Ca I lines at $\sim$6122 \AA~and 6573 \AA~generally used
for this determination in other studies  are unfortunately not sensitive enough
to metallicity nor to surface  gravity at the very low temperature of
our sample stars (see figure 7 of Cenarro et al.\ 2002). This was checked
by running different test models and spectral syntheses in the 6100--6160
\AA~and 6535--6585 \AA~regions where these lines fall. A decrease in the
metallicity implies less availability of metals like Ti, and thus, lowers the
TiO veiling, making the detection of atomic lines easier. Actually, this may be
the main reason why they are easily  detected in O-rich AGB stars at the
metallicity of the MCs (e.g.\ \textit{Z} = $-$0.5 in the SMC, Plez, Smith \&
Lambert 1993), while we do not detect them in our spectra.

\subsubsection{Zirconium}
The synthetic spectra are also very sensitive to changes in the zirconium
elemental abundance, especially in those wavelength regions where ZrO molecular
bands are present. For a set of fixed stellar parameters, an increase in the
zirconium abundance leads to stronger bands. Since the zirconium  abundance was
a priori unknown in our stars, we decided to use a large set of zirconium
abundances with {log $\varepsilon$(Zr)} between 1.6 and 3.6 dex in steps
of 0.25 dex.  The effect of a variation in the  zirconium elemental abundance
on the ZrO bands is in general much stronger  than any other effect associated
with variations in the {$T_{\rm eff}$}  or {C/O} ratio. In contrast,
even at relatively high zirconium abundance, the corresponding Zr atomic lines
are more difficult to detect at low {$T_{\rm eff}$} or low {C/O}
ratio because the TiO and ZrO lines become stronger and more numerous under
these conditions compared to the Zr atomic lines.

\subsubsection{CNO abundances and {$^{12}$C/$^{13}$C} ratio}
In the synthetic spectra, relative variations in the carbon, nitrogen and 
oxygen abundances and of the {$^{12}$C/$^{13}$C} ratio are completely
negligible with respect to other stellar fundamental parameters in the spectral
regions under analysis. Thus, we decided to fix them and adopt their solar
values in the following analysis.

\subsection{Abundance determination}
\subsubsection{Overall strategy}
After imposing these further constraints, the grid of MARCS model spectra  to
consider is composed of ``only'' a few hundred spectra, with effective
temperatures in the range {$T_{\rm eff}$} = 2500--3800 K in steps of 100 K,
{log $\varepsilon$(Zr)} between 1.6 and 3.6 dex in steps of 0.25 dex,
and a variable value of the {FWHM} in the range 200--600 m\AA~in steps
of 50 m\AA,  keeping all the other stellar parameters fixed: {log
$g$} = $-$0.5, $M$ = 2 $M_\odot$, solar metallicity {$ Z= $[Fe/H]}=0.0,
 {$\xi$} = 3 km s$^{-1}$,  {C/O} = 0.5, solar CNO abundances and solar
 {$^{12}$C/$^{13}$C} ratios.

In order to find the synthetic spectrum which better fits the observed spectrum
of a given star a modified version of the standard  $\chi$$^{2}$ test was
used.  When fitting observed data \textit{Yobs$_{i}$} to model data
\textit{Ysynth$_{i}$}, the quality of the fit can be quantified by the 
$\chi$$^{2}$ test. The best fit corresponds to that leading to the  minimum
value of $\chi$$^{2}$. This test is used here in a modified way and is 
mathematically expressed as:  

\begin{equation} 
\chi^2 = \sum_{i=1}^{N} \frac{ \left[
Yobs_{i}-Ysynth_{i} \left( x_{1}...x_{M} \right) 
\right]^2}{Yobs_{i}}  
\end{equation}   

with $N$ the number of data points, and $M$ the number of free parameters.  

The observed spectra, once fully reduced, were shifted to rest-wavelength using
the mean radial velocity shift derived from the Li I and Ca I lines. In
addition, they were re-binned to the same resolution as the synthetic ones
(0.06 \AA/pix) and normalized in order to make the comparison easier. The
observed spectra were then compared to the synthetic ones in intervals of
$\sim$60 \AA~which allowed the analysis of their overall characteristics as
well as of the  relative strength of the TiO molecular bands. 

The fitting procedure makes a special emphasis on the goodness of the fit in
the lithium region (6670--6730 \AA). In this spectral range the goal was to
fit the relative strength of the TiO bandheads at $\sim$6681 and 6714 \AA,
which are very sensitive to the effective temperature, together with the
pseudo-continuum around the Li I line. We first determined by $\chi$$^{2}$
minimization which of the spectra from our grid of models provided the  best
fit, mainly   fixing {$T_{\rm eff}$}. The best fits resulting from the
$\chi$$^{2}$ minimization were also judged by eye in order to test the method.
The lithium content was then estimated by changing the lithium abundance. This
procedure was repeated on each star of the sample for which an acceptable S/N
ratio was achieved around the Li I 6708 \AA~line. Unfortunately, we could not
analyse a few AGB stars with a low signal-to-noise ratio in their spectra at
6708 \AA~ (see Table 7). IRAS 18025$-$2113, IRAS 03507$+$1115 and IRAS
18304$-$0728 showed unusual spectra, very different from the rest of the stars
in our sample and a good fit from our grid of MARCS model spectra was not
possible. The spectra of these stars show unusual TiO band strengths which
could only be reproduced with a low {$T_{\rm eff}$} but the rest of the
spectrum (e.g.\ the local continuum  level) seems to be hotter. In addition, the
atomic lines are broader (IRAS 18025$-$2113 and IRAS 18025$-$2113) or narrower
(IRAS 03507$+$1115) than the {FWHM} needed to describe the TiO bandheads
and lines. Their complicated spectra look like a combination of two
temperatures and {FWHMs} suggesting that they could probably be 
double-lined spectroscopic binaries or strongly affected by shock waves propagating
in  their atmosphere. 

\begin{figure*}
\centering
\includegraphics[width=9cm,height=15cm,angle=-90]{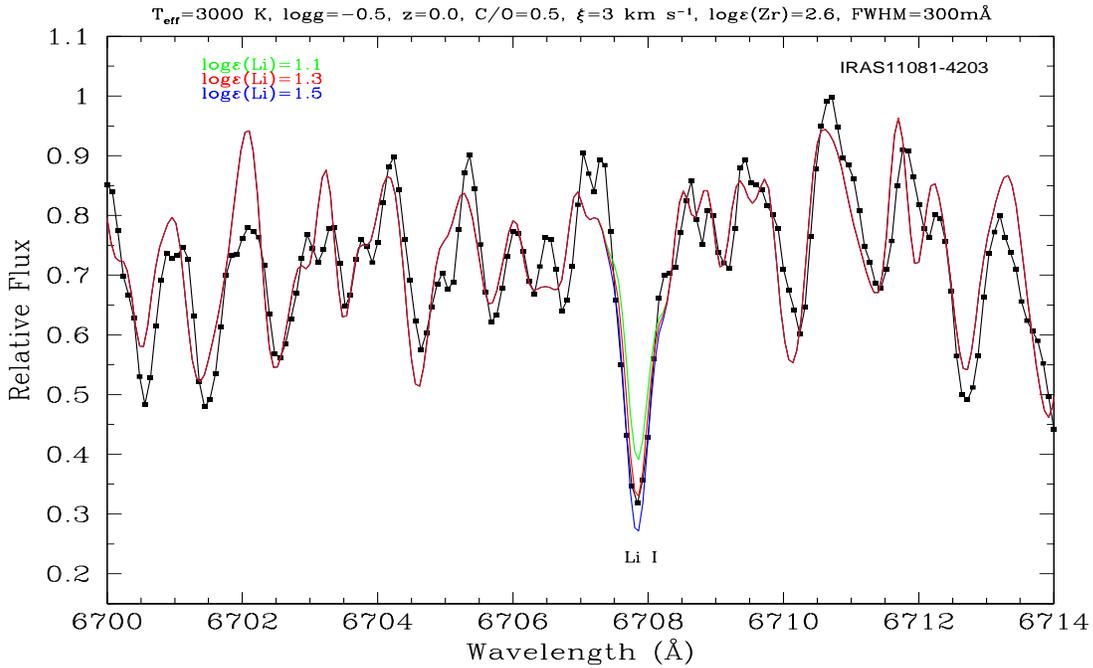}
\caption{Best model fit and observed spectrum around the Li I line (6708 \AA)
for the star IRAS 11081$-$4203. The Li abundance derived from this spectrum was
{log $\varepsilon$(Li)} = 1.3 dex. The synthetic spectra obtained for Li
abundances shifted $+$0.2 dex and $-$0.2 dex from the adopted value are also
shown. The parameters of the best model atmosphere fit are indicated in the top
label.}
\end{figure*}

\begin{figure*}
\centering
\includegraphics[width=9cm,height=15cm,angle=-90]{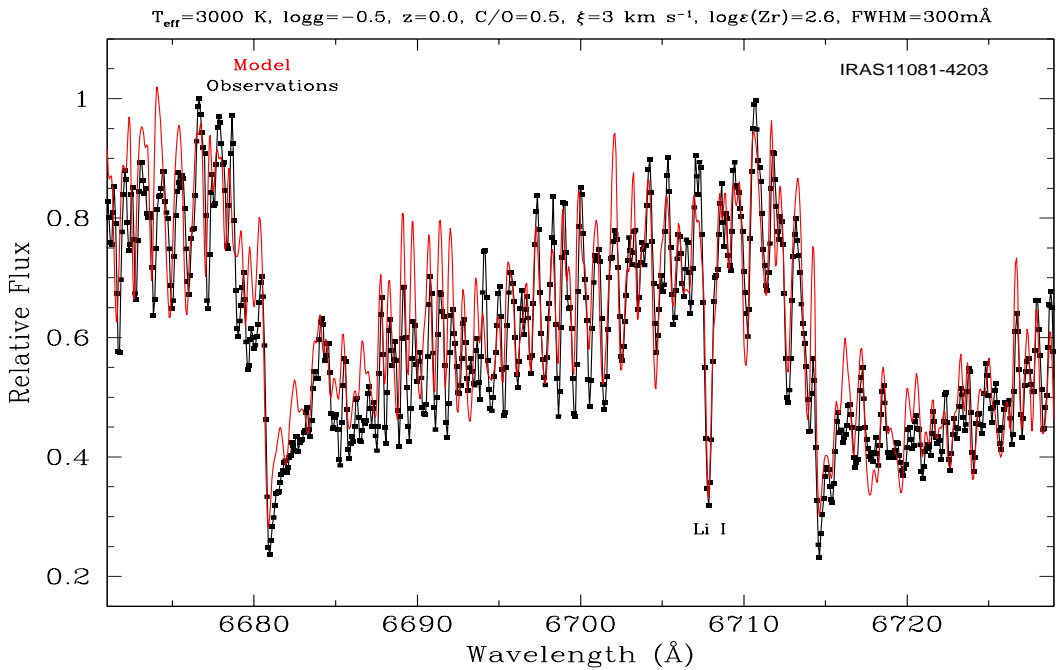}
\caption{Best model fit and observed spectrum in the region 6670--6730 \AA~for
the star IRAS 11081$-$4203. The {$T_{\rm eff}$} derived from this spectrum 
was 3000 K. The parameters of the best model atmosphere fit are indicated 
in the top label.}
\end{figure*}

\begin{figure*}
\centering
\includegraphics[width=9cm,height=15cm,angle=-90]{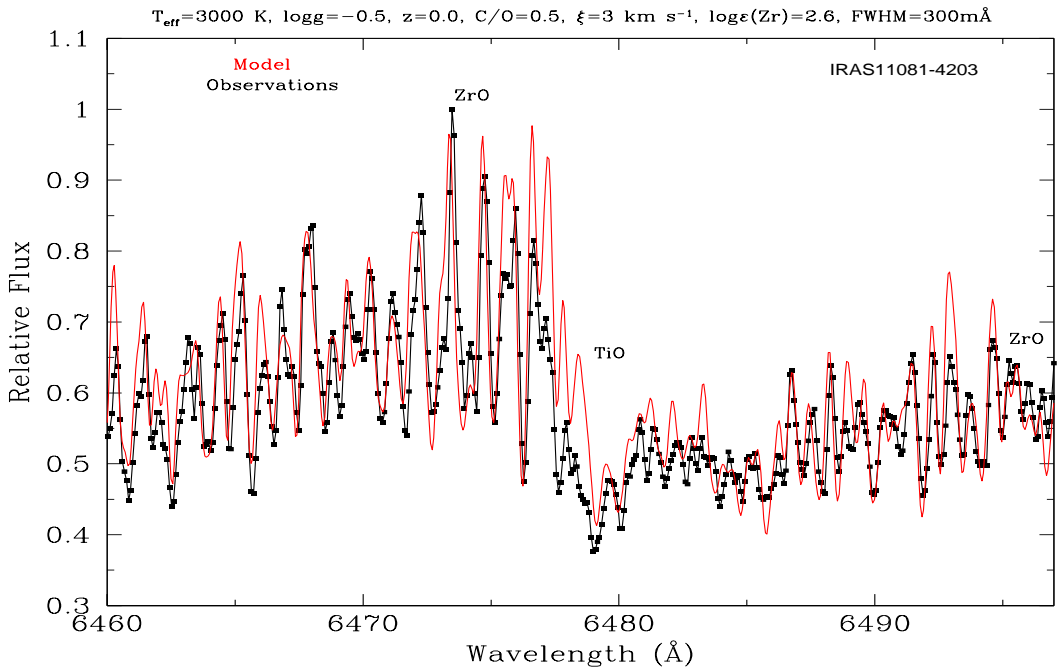}
\caption{Best model fit and observed spectrum in the region 6460--6499
\AA~for the star IRAS 11081$-$4203. The Zr abundance derived from this 
spectrum was {log $\varepsilon$(Zr)} = 2.6 dex, which corresponds to the 
solar value. The parameters of the best model atmosphere fit are indicated in 
the top label. Note the non-detection of the characteristic strong ZrO 
bandheads typical of Galactic S-stars.}
\end{figure*}

\begin{figure*}
\centering
\includegraphics[width=9cm,height=15cm,angle=-90]{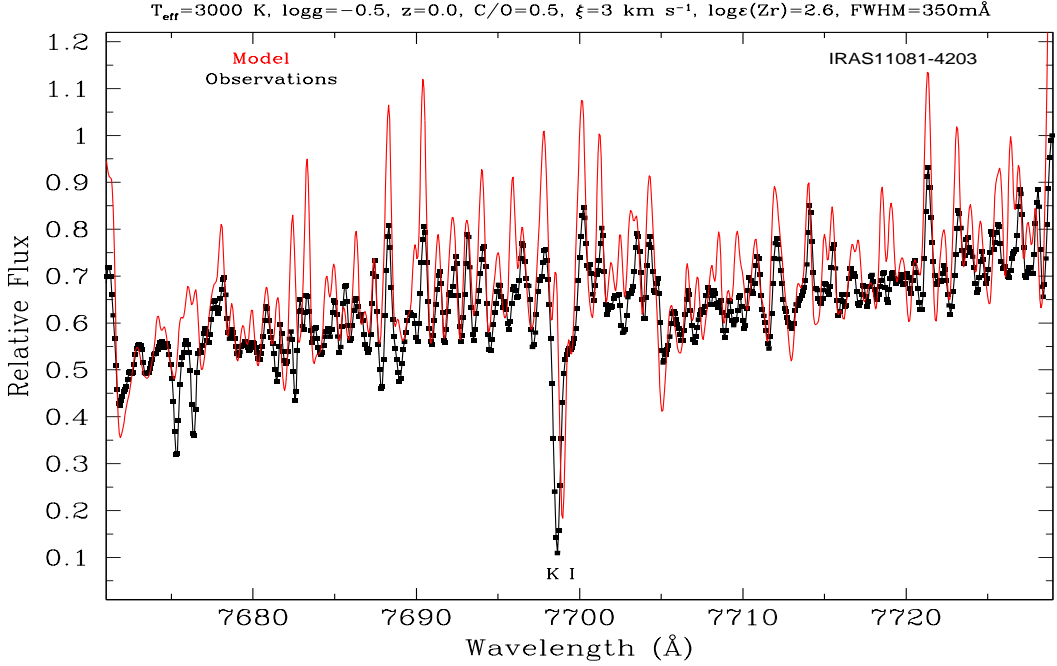}
\caption{Best model fit and observed spectrum in the region 7670--7730
\AA~(around the K I line at 7699 \AA) for the star IRAS 11081$-$4203. The 
synthetic spectrum obtained for solar K abundance ({log
$\varepsilon$(K)} = 5.12 dex) is shown. The parameters of the best model
atmosphere fit are indicated in the top label. Note the radial velocity
 shift, which is interpreted as a result of the partially 
 circumstellar origin of the
 K I line. The narrow features seen at $\sim$7675--77 \AA, which are not 
 adjusted by the model, are telluric lines.}
\end{figure*}

\begin{figure*}
\centering
\includegraphics[width=9cm,height=15cm,angle=-90]{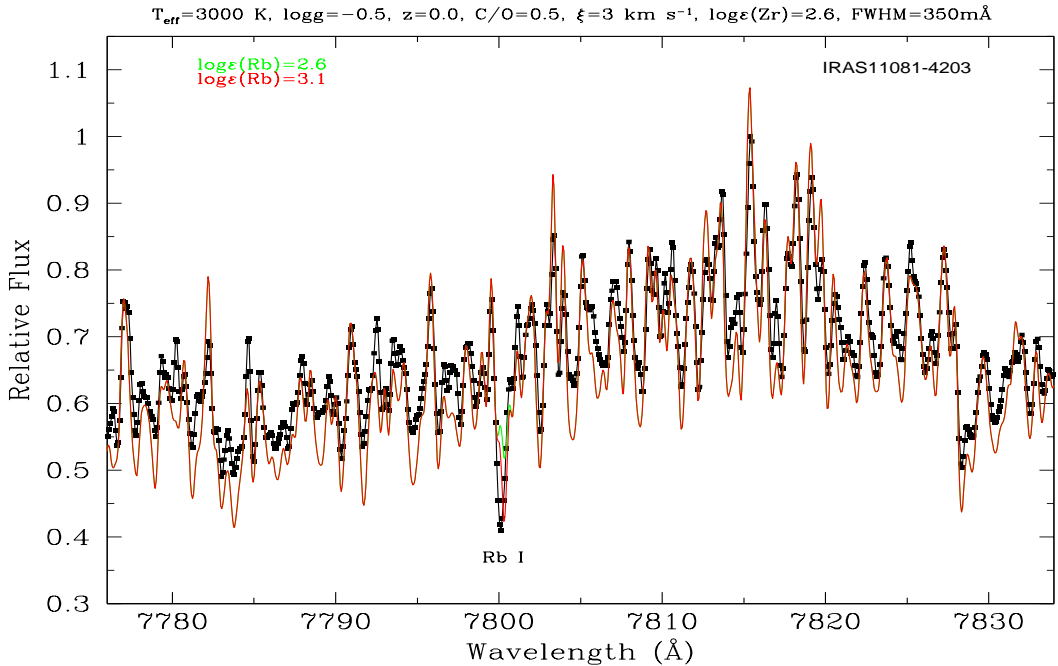}
\caption{Best model fit and observed spectrum in the region
7775--7835\AA~(around the Rb I line at 7800 \AA) for the star IRAS
11081$-$4203. The synthetic spectra obtained for solar Rb abundance
({log $\varepsilon$(Rb)} = 2.6 dex) and shifted $+$0.5 dex 
({log $\varepsilon$(Rb)} = 3.1 dex) are shown. The parameters of the
best model atmosphere fit are indicated in the top label. Again, the radial
velocity shift observed suggests a partially circumstellar origin of 
the Rb I line.}
\end{figure*}

\begin{figure*}
\centering
\includegraphics[width=9cm,height=15cm,angle=-90]{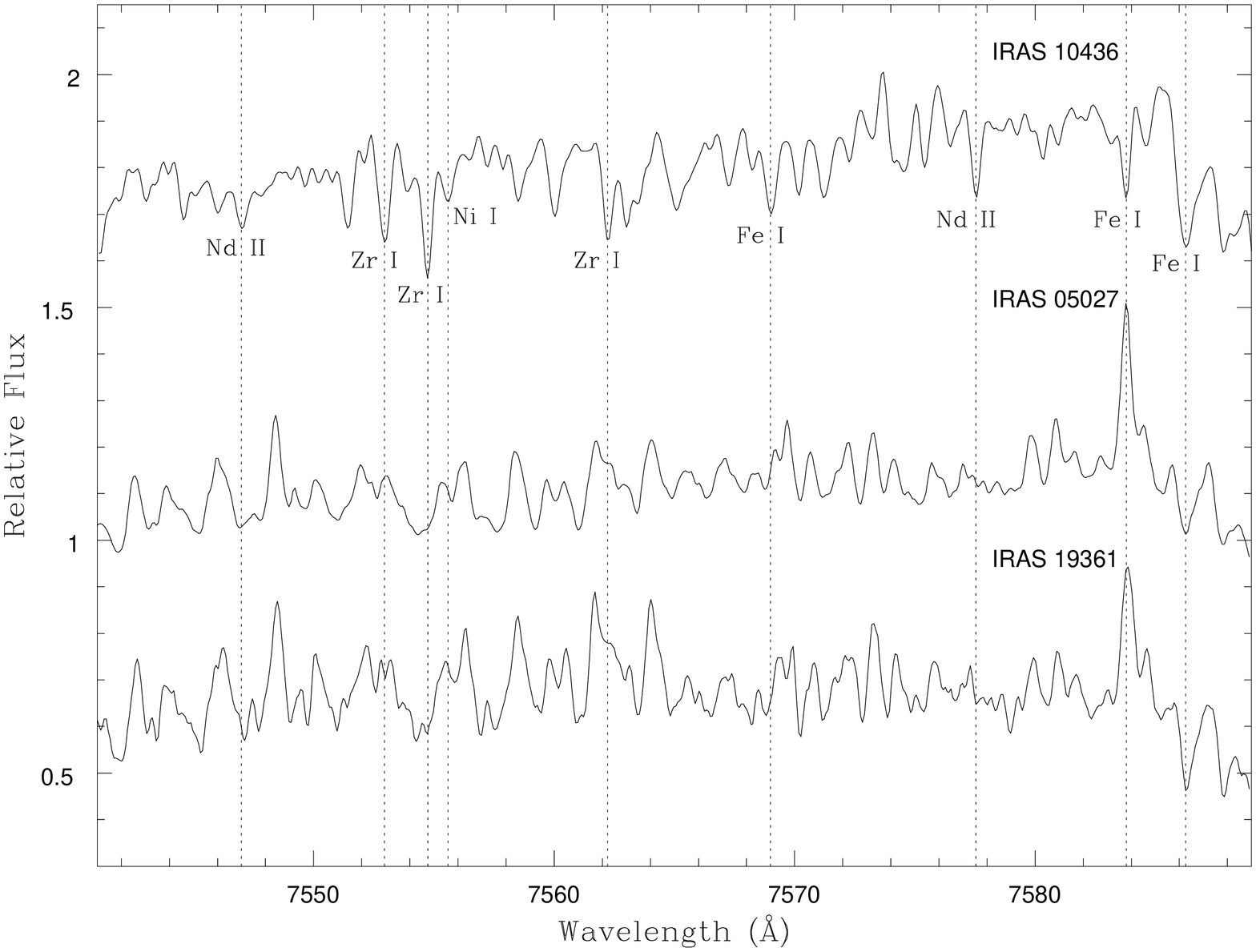}
\caption{High resolution optical spectra around the 7565 \AA~region of two
stars in our sample where we show the lack of s-process element
atomic lines (e.g.\ from Zr and Nd)  in comparison with the Galactic S-star IRAS
10436$-$3459. IRAS 05027$-$2158 and IRAS 19361$-$1658 show lithium lines with
very different strength. The positions of some atomic lines of Zr, Nd and Fe
are indicated. It seems that only the Fe I line at $\sim$7586 \AA~is clearly
detected in all stars.}
\end{figure*}

The best model fit in the Li I spectral region (6670--6730 \AA) is also  usually
found to fit the wavelength regions around the ZrO
bandhead and the K I and Rb I resonance lines reasonably well. In general, the overall shape of
the spectrum (including the TiO bandheads) is very well reproduced. As an
example, the fits made in different spectral regions for the star IRAS
11081$-$4203 are presented in Figures 5 to 9. The effective temperatures of the
best-fitting model spectra together with the lithium abundances (or upper
limits) derived following the above procedure are listed in Table 10, where we
have separated the Li detected stars from the Li undetected ones.

For the measurement of the zirconium abundance we concentrated our attention
on the synthesis of the 6455--6499 \AA~spectral region. In this  case, the
goal was to obtain the best fit around the ZrO molecular bands     at 6474
\AA~and 6495 \AA. The non-detection of these features in any of the   stars
under analysis imposes severe upper limits to the zirconium abundance.  To be
sure that there was no other way to interpret our data, we also checked the
presence of atomic zirconium in the 7400--7600 and 8100--8150 \AA~spectral
regions, where strong Zr I atomic lines should have been detected in that case.
Unfortunately, the local continuum in these latter wavelength regions is not
well reproduced in the O-rich AGB  stars in our sample, in contrast with the
perfect agreement obtained in some  of the comparison stars observed,  such as the
S-star IRAS 10436$-$3454, for which  a perfect fit is obtained both for the ZrO
bandheads at 6455$-$6499 \AA~ (see Figure 12) as well as for the Zr I atomic
lines in the 7400--7600 and  8100--8150 \AA~spectral regions (not shown). Sample
spectra of the region  around 7565 \AA~are shown in Figure 10, where the
position of some atomic  lines of Zr I, Nd II and Fe I are indicated.  As we
can see, the atomic lines corresponding to s-process elements (e.g.\ Zr, Nd, La,
etc.) were not detected in any of the sample stars observed. In contrast,
these  s-element atomic lines appear very strong in the Galactic S-star  IRAS
10436$-$3454.   Note, however, that the stars in our sample are the coolest
O-rich AGB stars ever studied (with effective temperatures between 2700 K and
3300 K; see Table 10).  We suspect that the  strong discrepancies observed
between model and observations may be due to the non-inclusion of other O-rich
molecules, such as  H$_{2}$O, in our line lists. The  effect of H$_{2}$O may
be ignored in S-stars, where the {C/O} ratio is close to unity, but it
might become dominant in our O-rich stars at these very  low temperatures,
preferentially at the longer wavelengths (e.g. Allard, Hauschildt \&
Schwenke 2000; Decin et al.\ 2003a, 2003b). In  addition, lanthanum oxide also has 
a strong absorption band around 7500 \AA~and it could have been a good 
candidate for detection in the spectral range here considered. As for ZrO, the
non-detection of the strong LaO molecular bandhead,  usually very strong in
S-stars at these wavelengths,  is another indication of the lack of s-process
elements in our sample stars.

\subsubsection{Lithium and s-element abundances}
\begin{table*}
\begin{center}
\caption[]{Spectroscopic $T_{\rm eff}$ and Li abundances$^{1}$ derived for the
subgroup of Li detected (left) and Li non-detected stars (right)}
\footnotesize
\begin{tabular}{lcccclcc}
\hline\hline
\multicolumn{3}{c}{Li detected}& ~~~~~~~~~~& 
\multicolumn{3}{c}{Li non-detected}\\
\hline
& \textit{T$_{eff}$} & \textit{log $\varepsilon$(Li)}& &
 &\textit{T$_{eff}$}& \textit{log $\varepsilon$(Li)}\\
IRAS name&K& 12 $+$ {log $N$(Li)} & &IRAS name &K &12 $+$ {log $N$(Li)}\\ 
\hline
01085$+$3022 & 3300 & 2.4  	    &&03507$+$1115$^{2}$ &$\dots$&$\dots$\\
02095$-$2355 & 3300 & 1.6          &&05098$-$6422 &3000&$<$$-$1.0\\
04404$-$7427$^{**}$ &3000&$\dots$ &&05151$+$6312 &3000&$<$+0.0\\
05027$-$2158 & 2800 & 1.1            &&07080$-$5948 &3000&$<$+0.5\\
05559$+$3825 & 2900 & 0.6          &&07445$-$2613 &2900&$<$$-$1.0\\
06300$+$6058 & 3000 & 0.7       &&10189$-$3432 &2900&$<$$-$1.0\\
07222$-$2005$^{**}$ &3000 &$\dots$ &&10261$-$5055 &3000&$<$$-$1.0\\
07304$-$2032 & 2700 & 0.9         &&13379$-$5426 &2900&$<$+0.0\\
09429$-$2148 & 3300 & 2.2         &&13442$-$6109$^{4}$ &3000&$<$$-$1.0\\
11081$-$4203 & 3000 & 1.3        &&13475$-$4531 &2900&$<$$-$1.0 \\
11525$-$5057 & 3300&  0.9 &&14086$-$0730$^{**}$&$\dots$&$\dots$\\
12377$-$6102$^{**}$ &$\dots$&$\dots$ &&14086$-$6907$^{**}$ &$\dots$&$\dots$\\
14337$-$6215 & 3300 &2.4$^{*}$   &&14247$+$0454 &2700&$<$+0.5\\
15211$-$4254 & 3300 & 2.3       &&14266$-$4211 &2900&$<$+0.0\\
15255$+$1944 & 2900 & 1.0      &&15193$+$3132 &2800&$<$+0.0\\
15576$-$1212 & 3000 & 1.1 	    &&15586$-$3838 &3000&$<$+0.0\\
16030$-$5156 & 3000 &1.5     &&16037$+$4218 &2900&$<$$-$1.0\\
16260$+$3454 & 3300 & 2.7	    &&16503$+$0529 &2800&$<$$-$1.0\\
18025$-$2113$^{2}$&$\dots$&$\dots$  &&17034$-$1024 &3300&$<$+0.0\\
18057$-$2616$^{3}$ & 3000 &$\dots$ &&17359$-$2138$^{**}$ &$\dots$&$\dots$\\
18413$+$1354 & 3300 &1.8    &&18050$-$2213 &2900&$<$$-$1.0\\
18429$-$1721 & 3000 &1.2     &&18304$-$0728$^{2}$ &$\dots$&$\dots$\\
19129$+$2803 & 3300 &3.1$^{*}$ &&18454$-$1226$^{4}$&3300&$<$+0.0\\
19361$-$1658$^{5}$&3000&1.9/2.0 &&19147$+$5004 &3000&$<$+0.0\\
20052$+$0554 & 3300 &2.6   &&19157$-$1706 &3300&$<$+0.0\\
 & &                                &&19412$+$0337 &3300&$<$+0.0\\    
 & &                   	            &&20343$-$3020 &3000&$<$$-$1.0\\ 
 & &                   		    &&20350$+$3741  &3000&$<$+0.0\\   
 & &                   		    &&22180$+$3225 &3300&$<$+0.0\\    
 & & 			            &&22466$+$6942&3300&$<$+0.0\\
 & & 			            &&22510$+$3614 &3000&$<$$-$1.0&\\
 & & 				    &&23492$+$0846$^{4}$ &3000&$<$$-$1.0\\
\hline
\end{tabular}
\end{center}
$^{1}$ The uncertainty in the derived Li abundances is estimated to be 
around $\pm$0.4$-$0.6 dex\\
$^{2}$ Possible double lined spectroscopic binary\\
$^{3}$ The lithium line has a P cygni-type profile\\
$^{4}$ Possible non-AGB star\\
$^{5}$ Observed Li abundances in two different runs\\
$^{*}$ The abundance value must be treated with some caution because the line is
resolved in two components (circumstellar and stellar) and the abundance
estimate corresponds to the photospheric abundance needed to fit the stellar
component\\
$^{**}$ The S/N ratio at 6708 \AA~is very low to derive any reliable 
{$T_{\rm eff}$} and/or Li abundance estimate\\
\end{table*}

\begin{table*}
\begin{center}
\caption[]{Sensitivity of the derived Li abundances (in dex) to slight 
changes in the model atmosphere parameters for IRAS 15255$+$1944}
\footnotesize
\begin{tabular}{ccc}
\hline\hline
Adopted value & Change & Li abundance\\
\hline
{$T_{\rm eff}$} = 2900 K & $\Delta$ { $T_{\rm eff}$} =  $\pm$100 K &
$\Delta${log $\varepsilon$(Li)} = $\pm$0.3\\
{$Z$} = 0.0 & $\Delta Z = \pm$ 0.3 &$\Delta${log
$\varepsilon$(Li)} = $\pm$0.2\\
{$\xi$} = 3 km s$^{-1}$&$\Delta${$\xi$} = $\pm$1 km
s$^{-1}$&$\Delta${log $\varepsilon$(Li)} = $\mp$0.1\\
{log $g$} = $-$0.5 & $\Delta${log $g$} = $+$0.5 &$\Delta${log
$\varepsilon$(Li)} = $+$0.1\\
{FWHM} = 300 m\AA &$\Delta${FWHM} = $\pm$50 m\AA & 
$\Delta${log $\varepsilon$(Li)} = $\pm$0.1\\
\hline
\end{tabular}
\end{center}
\end{table*}

The derived Li abundances are displayed in Table 10, where we can see  that
almost all Li detected stars show enhanced Li abundances {log
$\varepsilon$(Li)} $\geq$ 1, i.e.\ larger than solar, but smaller  than those
found in the so-called ``super Li-rich'' AGB stars (with {log
$\varepsilon$(Li)} $>$ 3$-$4; e.g.\ Abia et al.\ 1991) in our Galaxy. A
very similar range of Li overabundances was found in the massive O-rich AGB
stars studied in the MCs (Plez, Smith \& Lambert 1993; Smith \& Lambert 1989,
1990a; Smith et al.\ 1995). The errors in the derived Li abundances  mainly
reflect the sensitivity of our models  to changes in the adopted stellar
parameters. In particular, they strongly depend on the uncertainties in the
determination  of the effective temperature ($\pm$100--200 K; larger for the
coolest stars) and metallicity ($\pm$0.3), while they are less sensitive to
the  microturbulent velocity ($\pm$1 km s$^{-1}$), surface gravity ($\pm$0.5)
and {FWHM} ($\pm$50 m\AA) uncertainties.  The effect of the uncertainty
in the location of the pseudo-continuum around  the Li I line on the measured
abundances is negligible compared with any  other change in the adopted stellar
parameters. If we consider all these  uncertainties as independent sources of
error, the resulting  Li abundances given in Table 10 are estimated to be
affected by errors of the order of 0.4--0.6 dex. As an example, the changes in
the derived Li abundance induced by slight variations of each of the
atmospheric parameters used in our modelling for IRAS 15255$+$1944 are
shown in Table 11.

Note, however, that the estimated errors do not reflect possible non-LTE
effects, dynamics of the stellar atmosphere, or errors in the model
atmospheres or in the molecular/atomic linelists themselves. In particular,
over-ionization and over-excitation of Li is predicted to occur under non-LTE
conditions both in C-rich and O-rich AGB stars (Kiselman \& Plez 1995; Pavlenko
1996) although the effect is expected to be more pronounced in metal-deficient
stars (i.e.\ MC AGBs). Use of LTE in  Li-rich AGB stars is likely to result in
underestimation of the Li abundance  (Kiselman \& Plez 1995; Abia, Pavlenko
\& Laverny 1999).

\begin{figure*}
\centering
\includegraphics[width=9cm,height=15cm,angle=-90]{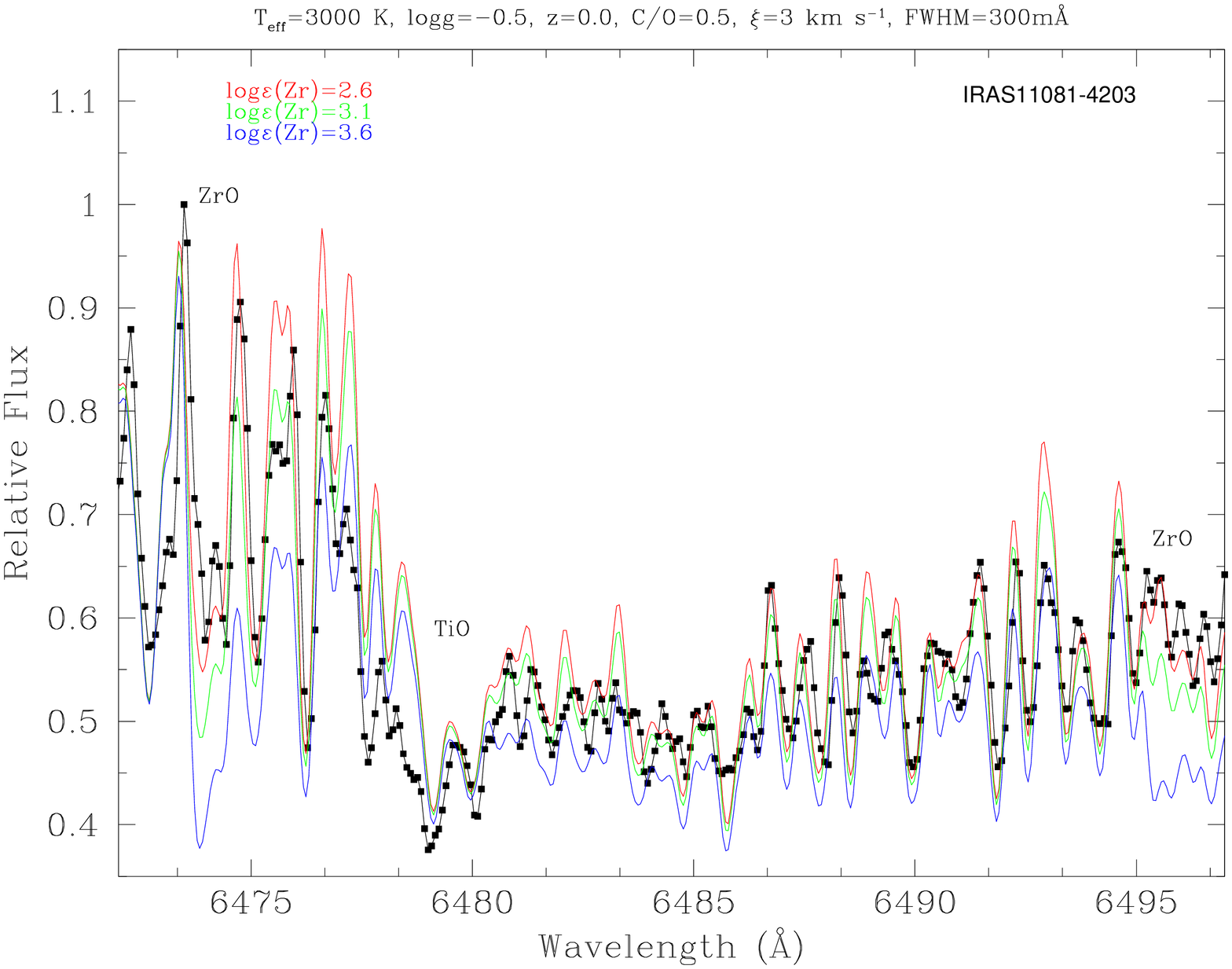}
\caption{Synthetic and observed spectra in the region 6460--6499 \AA~for the
star IRAS 11081$-$4203, one of the Li detected stars in our sample. The
synthetic spectra corresponding to [Zr/Fe] $=$ $+$0.0, $+$0.5, and $+$1.0 dex 
(or {log $\varepsilon$(Zr)} = 2.6, 3.1, and 3.6 dex, respectively) are 
shown. The parameters of the best model atmosphere fit are indicated in the top
label.}
\end{figure*}

\begin{figure*}
\centering
\includegraphics[width=9cm,height=15cm,angle=-90]{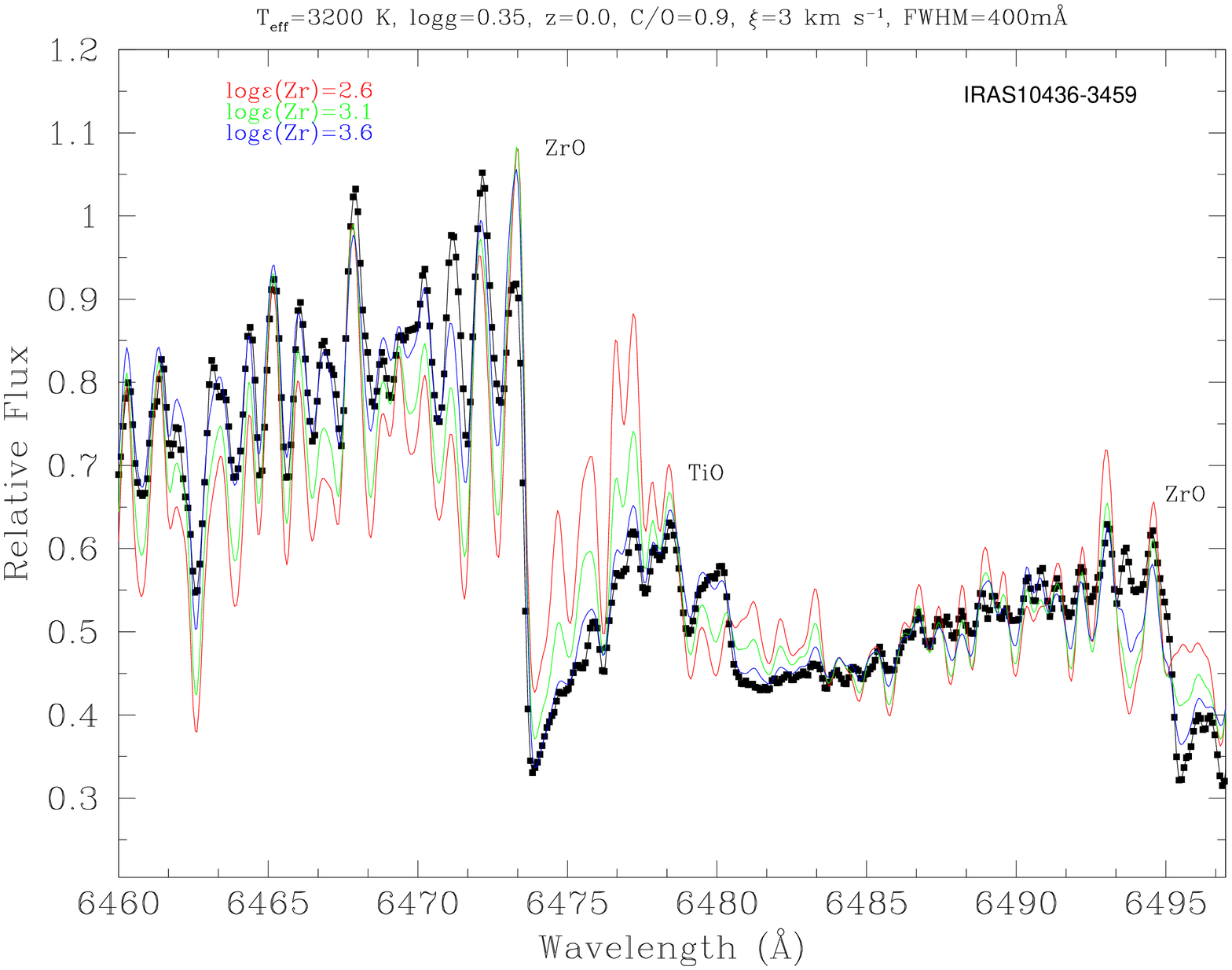}
\caption{Synthetic and observed spectra in the region 6460--6499 \AA~for the
Galactic S-star IRAS 10436$-$3454. The synthetic spectra corresponding to
[Zr/Fe] $=$ $+$0.0, $+$0.5  and $+$1.0 dex (or  {log
$\varepsilon$(Zr)} = 2.6, 3.1, and 3.6 dex, respectively) are shown. The
parameters of the best model atmosphere fit are indicated in the top label.
Note the strong overabundance of [Zr/Fe] $=$ $+$1.0 dex ({log
$\varepsilon$(Zr)} = 3.6 dex) needed to fit the observations.}
\end{figure*}

Concerning s-process elements, the non-detection of the ZrO molecular bands  at
6474 \AA~and 6495 \AA~is found to be consistent with upper limits for the  Zr
abundance around [Zr/Fe] $=$ 0.25 $\pm$ 0.25 dex with respect to the solar value of
{log $\varepsilon$(Zr)} = 2.6 dex. The synthetic spectra predict
detectable ZrO bandheads at 6474 \AA~and 6495 \AA~for Zr abundances above
0.00--0.25 dex for {$T_{\rm eff}$} $\geq$ 3000 K and above 0.25--0.50 dex
for  {$T_{\rm eff}$} $<$ 3000 K. Actually, the fits are also reasonably 
good even if we do not include any ZrO in the synthesis. We found the same
results both in the Li non-detected stars and in the Li detected ones. This
result is in strong contrast with the higher Zr abundances ([Zr/Fe] $>$ 0.5)
found in Galactic MS, S-stars and in massive O-rich AGB stars in the MCs. The
effect of varying the Zr abundance for the Li detected star IRAS 11081$-$4203
is displayed in Figure 11, as an  example. The best model spectrum fit for the
Galactic S-star IRAS 10436$-$3454 is also shown in Figure 12 for comparison.
As we can see, strong ZrO bandheads are  detected in IRAS 10436$-$3454, while
these are completely absent from the spectrum of IRAS 11081$-$4203. According
to our models, a modest Zr enhancement (with respect to the solar value) would
be enough for the ZrO molecular bands to show up in our sample stars despite
the strong veiling produced by the TiO molecule (see Figure 11). An even lower
Zr  abundance is enough to produce strong ZrO bands in S-stars, where the 
{C/O} ratio is $\sim$1 because of the weaker TiO veiling, as we can see
in the spectrum of IRAS 10436$-$3454 displayed in Figure 12.  This also favours
the visibility of the Zr I atomic lines in the 7400--7600 and 8100--8150
\AA~spectral regions in these stars, as  has already been mentioned.

\section{Discussion}
\subsection{Absolute luminosities}
In constrast to the studies made in the past on AGB stars in the Magellanic
Clouds, for which a common distance can be assumed leading to relatively well
determined absolute luminosities, the estimation of absolute luminosities for
the stars in our Galactic sample is very difficult, mainly because of the large
uncertainties involved in the determination of distances within our
Galaxy.\footnote{Note that, in addition, AGB stars are also strongly variable.
As an example, Engels et al.\ (1983) found an $M_{\rm bol}$ variation of 2 magnitudes
between the minimum and maximum light (from $-$5.0 to $-$7.1 mag respectively)
in the Galactic OH/IR star OH 32.8$-$0.3.} As a first approach, we can try to
estimate  luminosities using the  period-luminosity relationship
(for those sources with a well  determined period).  If the P--L relationship
for Galactic Mira variables found by Groenewegen \& Whitelock (1996) is applied
to our sample of OH/IR stars, we obtain $M_{\rm bol}$ around $\sim$$-$6 for those
stars with periods of 400--600 days while an $M_{\rm bol}$ of $\sim$$-$5.5 is
derived for periods around $\sim$300 days. Similar values were found among the
more massive AGB stars in the MCs. However, if we take those stars with the
longer periods in the sample ($\gtrsim$ 1000 days), we derive $M_{\rm bol}$
$\lesssim$ $-$7, which seems unrealistically high. This  suggests that the
period--luminosity relationship may not hold for the most extreme OH/IR stars
(Wood, Habing \& McGregor 1998). On the other hand, there is recent
observational evidence for the existence of low metallicity (e.g.\ in the LMC)
HBB AGB stars with luminosities brighter than the predictions of the core-mass
luminosity relation which have been atributed to an excess flux from HBB
(Whitelock et al.\ 2003). A similar effect could explain the  detection of AGB
stars with such high $M_{\rm bol}$ also in our Galaxy. 

\subsection{Progenitor masses}

 Neither is determining the progenitor masses of our Galactic O-rich AGB stars 
a simple task. Several observational parameters, however, such as the OH maser
expansion velocity or the variability period, have  been proposed in the past
as useful distance-independent mass indicators for this class of stars, and we
will make use of them  in the following. 

\begin{figure*}
\centering
\includegraphics[width=5.9cm]{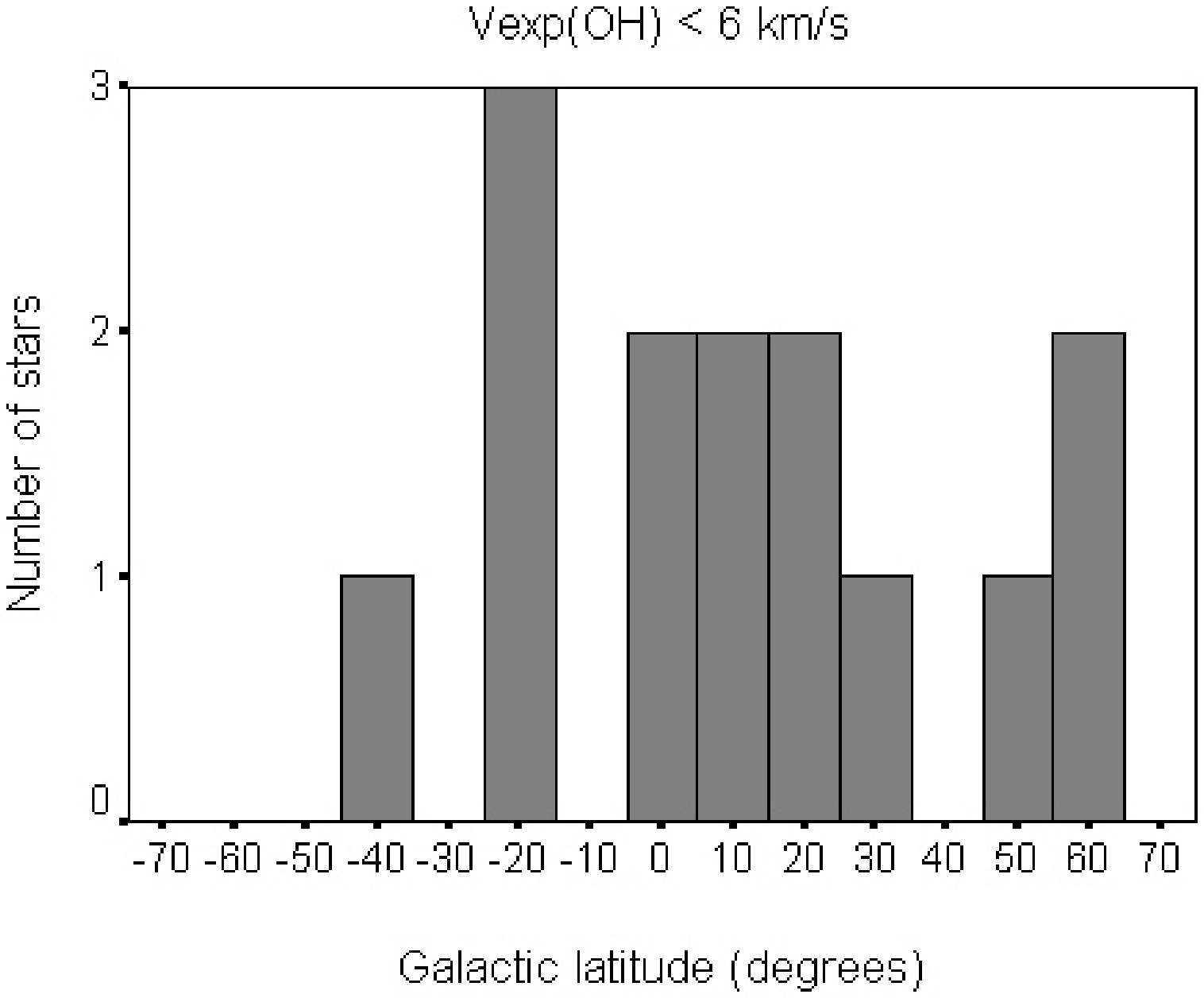}%
\includegraphics[width=5.9cm]{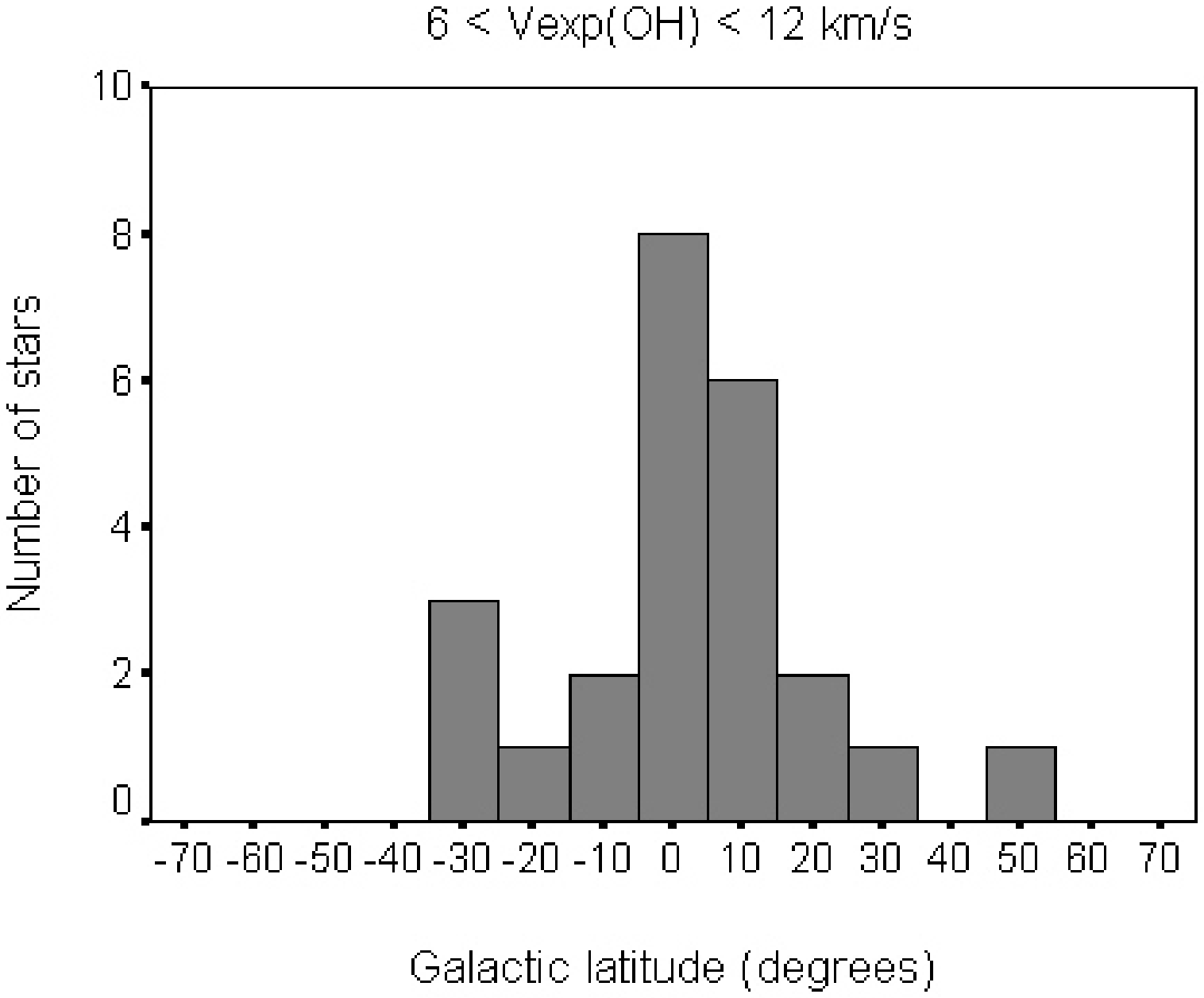}%
\includegraphics[width=5.9cm]{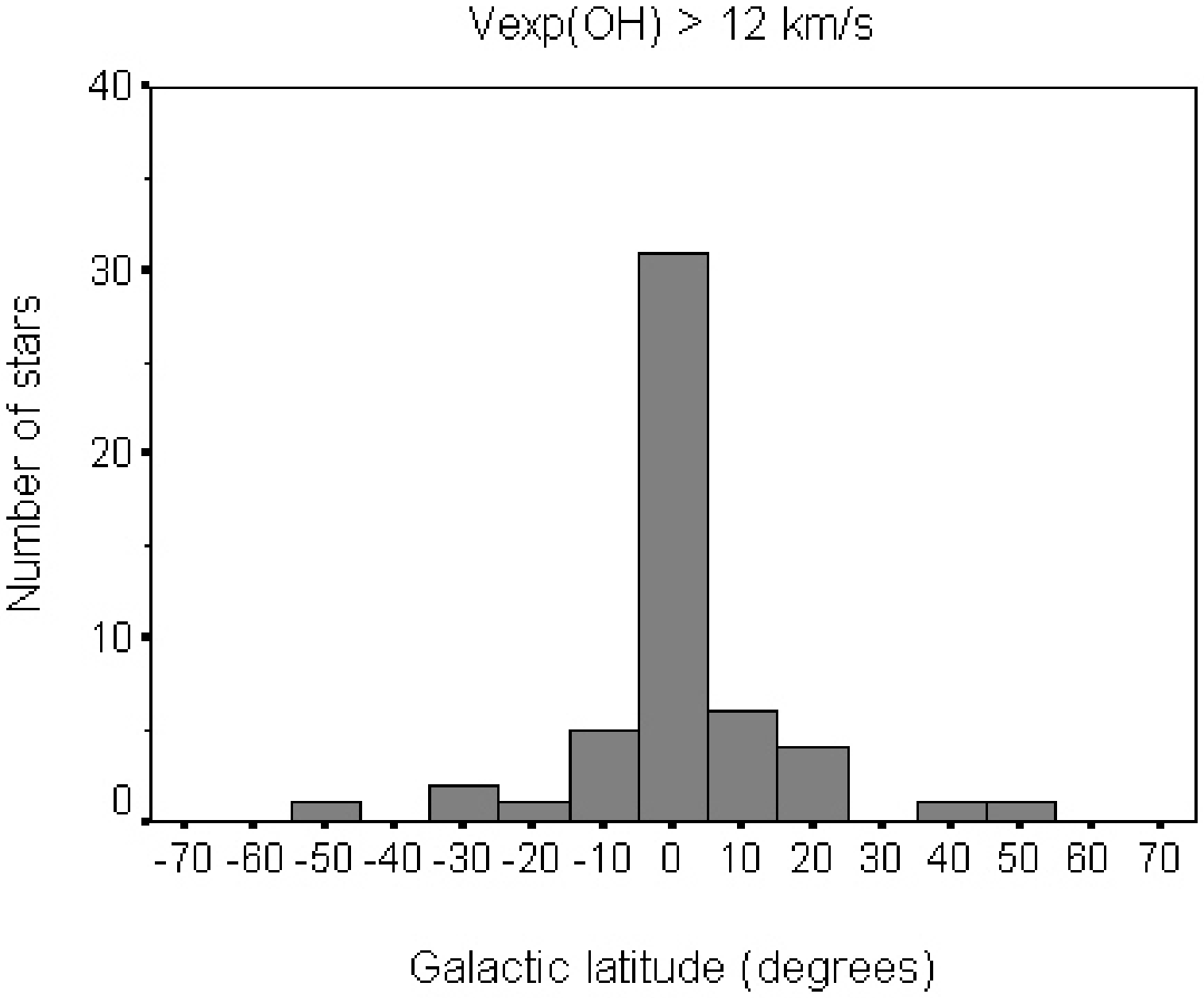}
\caption{Galactic latitude distribution of the sources
in our sample as a function of their OH expansion velocities: 
$v_{\rm exp}$(OH) $<$ 6 km s$^{-1}$ (left panel); 
6 $<$ $v_{\rm exp}$(OH) $<$ 12 km s$^{-1}$ (middle panel);
and $v_{\rm exp}$(OH) $>$ 12 km s$^{-1}$ (right panel).} 
\end{figure*}

The assumption is based on the fact that the group of OH/IR stars with  longer
periods and larger expansion velocities shows a Galactic distribution  that
corresponds to a more massive population (Baud et al.\ 1981;  Baud \& Habing
1983; Chen et al.\ 2001; Jim\'enez-Esteban et al.\ 2005b) suggesting that the 
periods of variability and the expansion velocities of the circumstellar 
envelopes of AGB stars must be closely correlated with the progenitor mass.

 Actually, stars in our sample follow the expected trend, as  is illustrated
in Figure 13. Although the limited number of stars  considered prevents us from
deriving any statistical conclusion based on our  data alone, it seems clear that
the galactic latitude  distribution of the stars in our sample becomes narrower
as a  function of $v_{\rm exp}$(OH), as one would expect if the higher expansion 
velocities do correspond to a more massive population. A similar trend is
observed when the variability period is analysed (not shown).

Figure 14 (left panel) shows the distribution of the sources in our sample on 
the IRAS two-colour diagram [12] $-$ [25] {\it vs} [25] $-$ [60] for which OH
expansion velocities are available. As we can see in this diagram, on average,
the sources with intermediate OH expansion velocities, between 6  and 12 km
s$^{-1}$, show redder [12] $-$ [25] and [25] $-$ [60] colours  than those with low
OH expansion velocities (i.e.\ below 6 km s$^{-1}$), which implies thicker
circumstellar envelopes or larger mass loss rates, but they are bluer than the 
subsample of stars with OH expansion velocities beyond 12 km s$^{-1}$. Redder
IRAS colours can be interpreted as the consequence of  a more advanced stage on
the AGB if we assume that mass loss rate increases along this evolutionary
phase. Alternatively, they may also be interpreted as corresponding  to a more
massive population of AGB stars since in principle they should be able to
develop thicker circumstellar envelopes  even at a relatively early stage in
their AGB evolution. A similar behaviour is observed when the variability
period is analysed (Figure 14; right panel), although the number of stars with
known  periods is significantly smaller. In this case we can see that the 
reddest sources in the diagram belong to the subgroup showing the longer
periods, in excess of 700 days.

\begin{figure*}
\centering
\includegraphics[width=6.8cm]{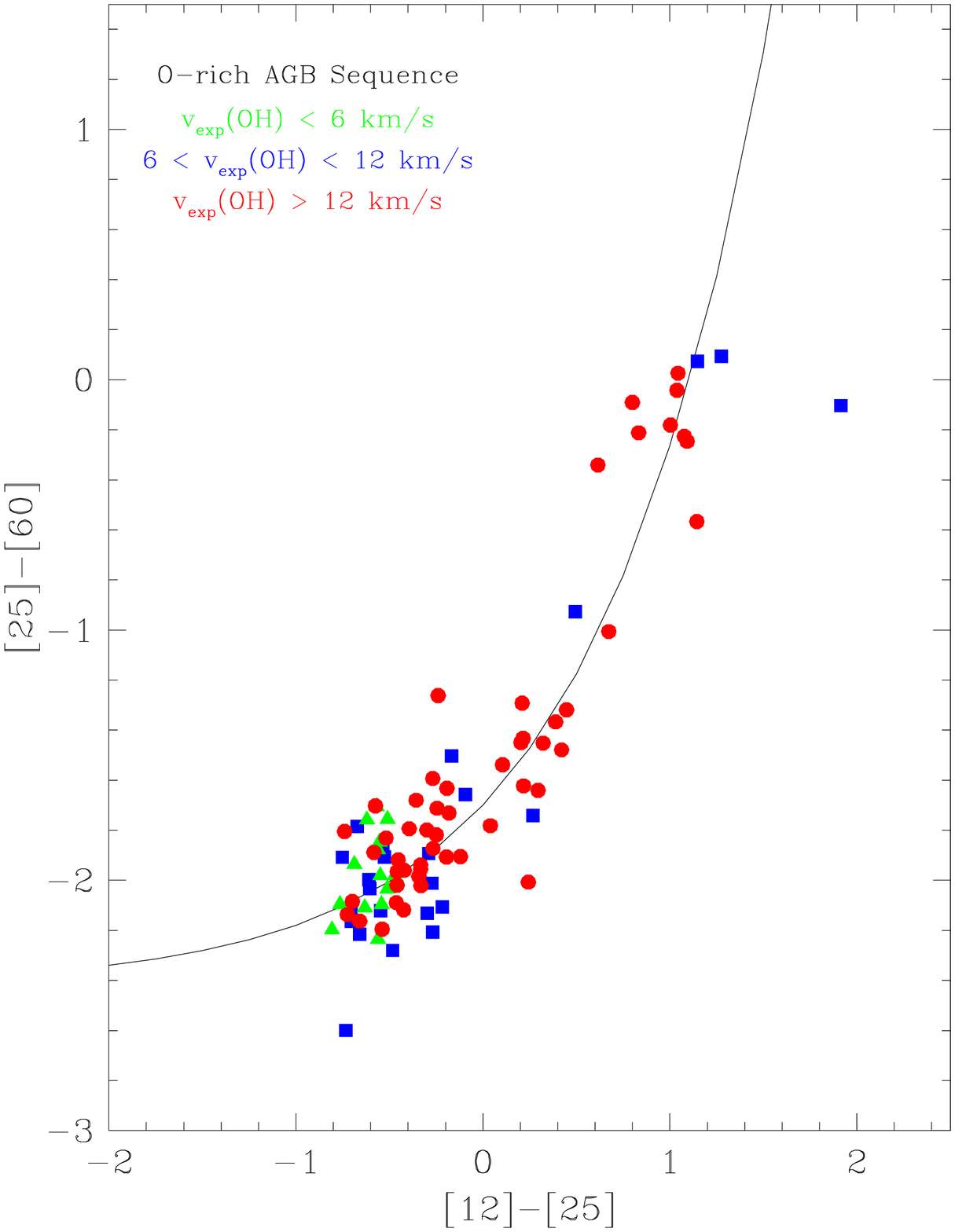}%
\includegraphics[width=6.8cm]{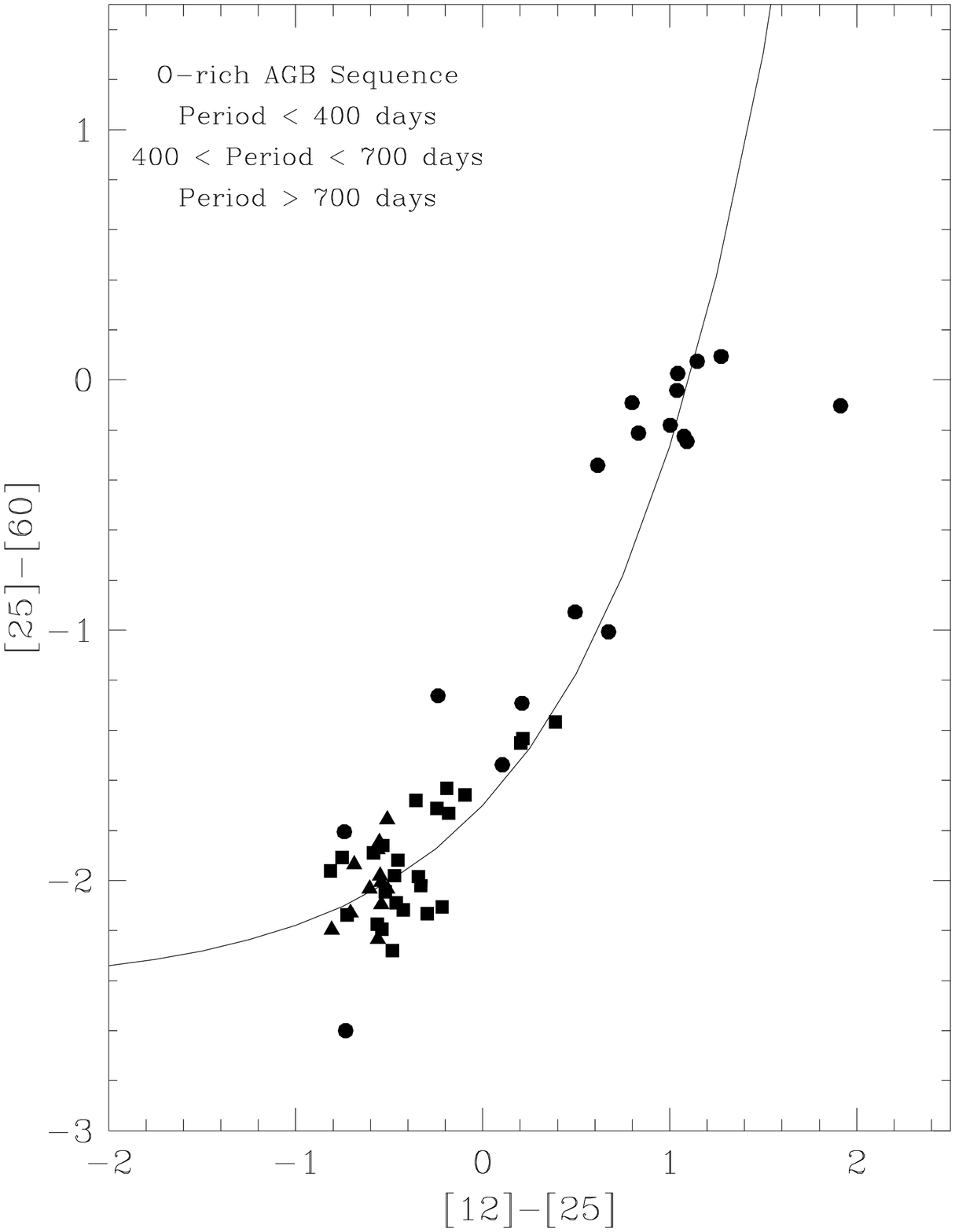}
\caption{Distribution of the sources included in our sample in the
IRAS two-colour diagram [12] $-$ [25] {\it{vs}} [25] $-$ [60] showing the location
of the stars in our sample as a function of the OH expansion velocities 
(left panel) and variability periods (right panel). The continuum
line is the ``O-rich AGB sequence'' (see text), which indicates the sequence of
colours expected for O-rich AGB stars surrounded by envelopes with increasing
thickness and/or mass loss rates (Bedijn 1987).}
\end{figure*}

The distribution of variability periods and OH expansion velocities is found to
be remarkably  different for each of the three subgroups identified in our 
sample. Histograms showing the results obtained are shown in Figures 15 and
16, respectively. The same results are presented in a different way in Table
12, where the distribution of sources among the different subtypes is presented
as a function of their OH expansion velocity and variability  properties.

\begin{figure*}
\centering
\includegraphics[width=6.1cm]{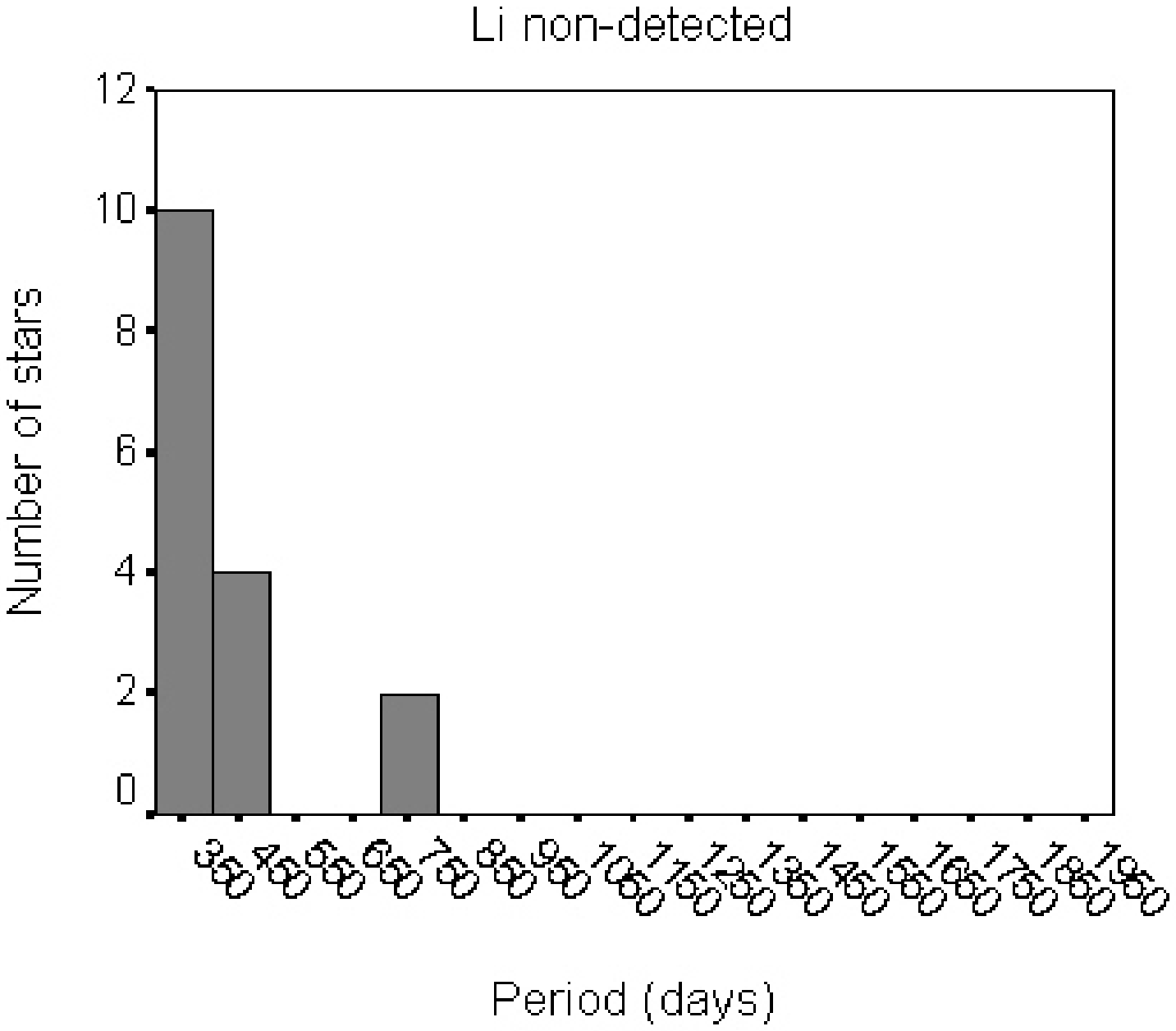}%
\includegraphics[width=6.1cm]{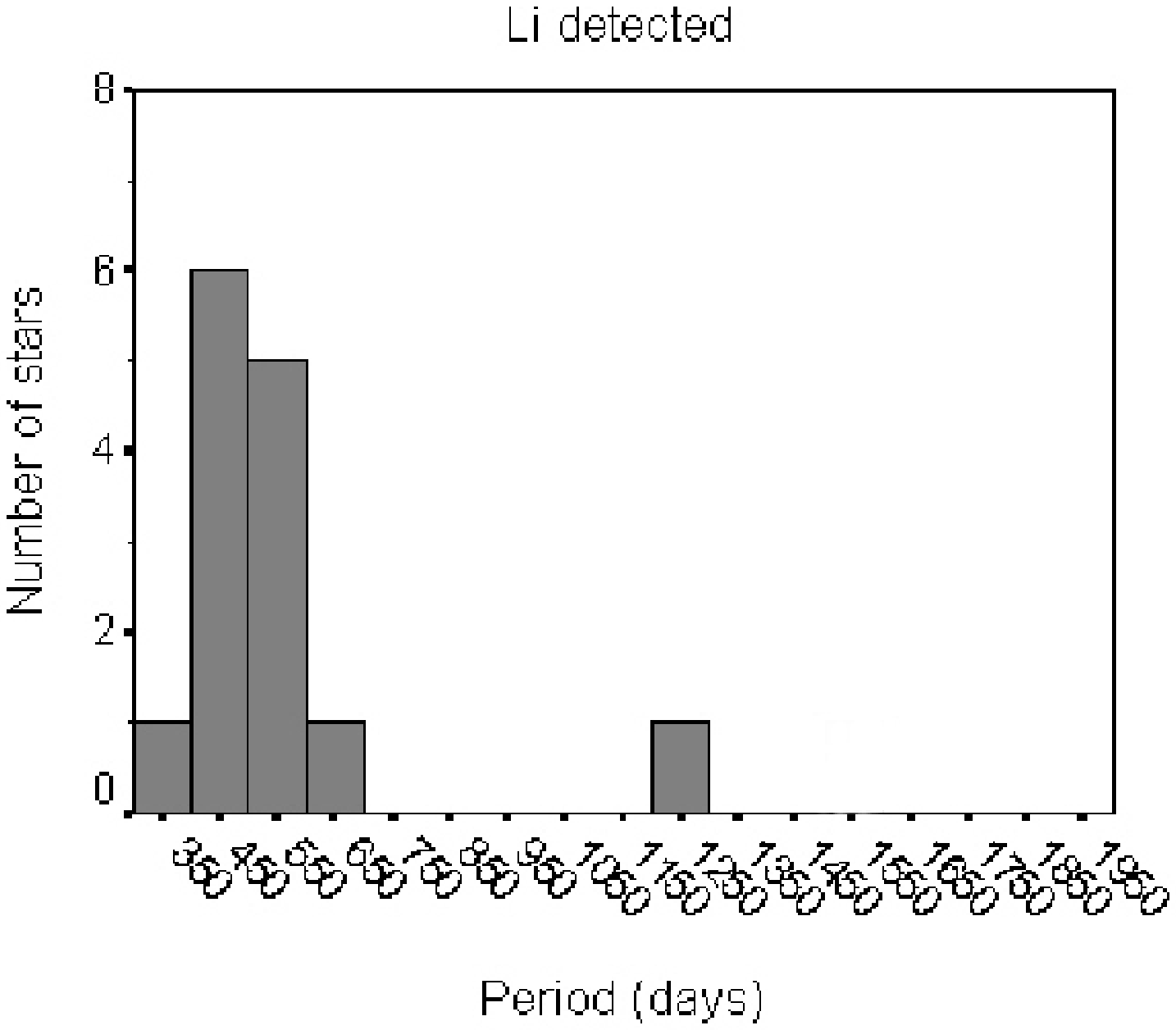}%
\includegraphics[width=6.1cm]{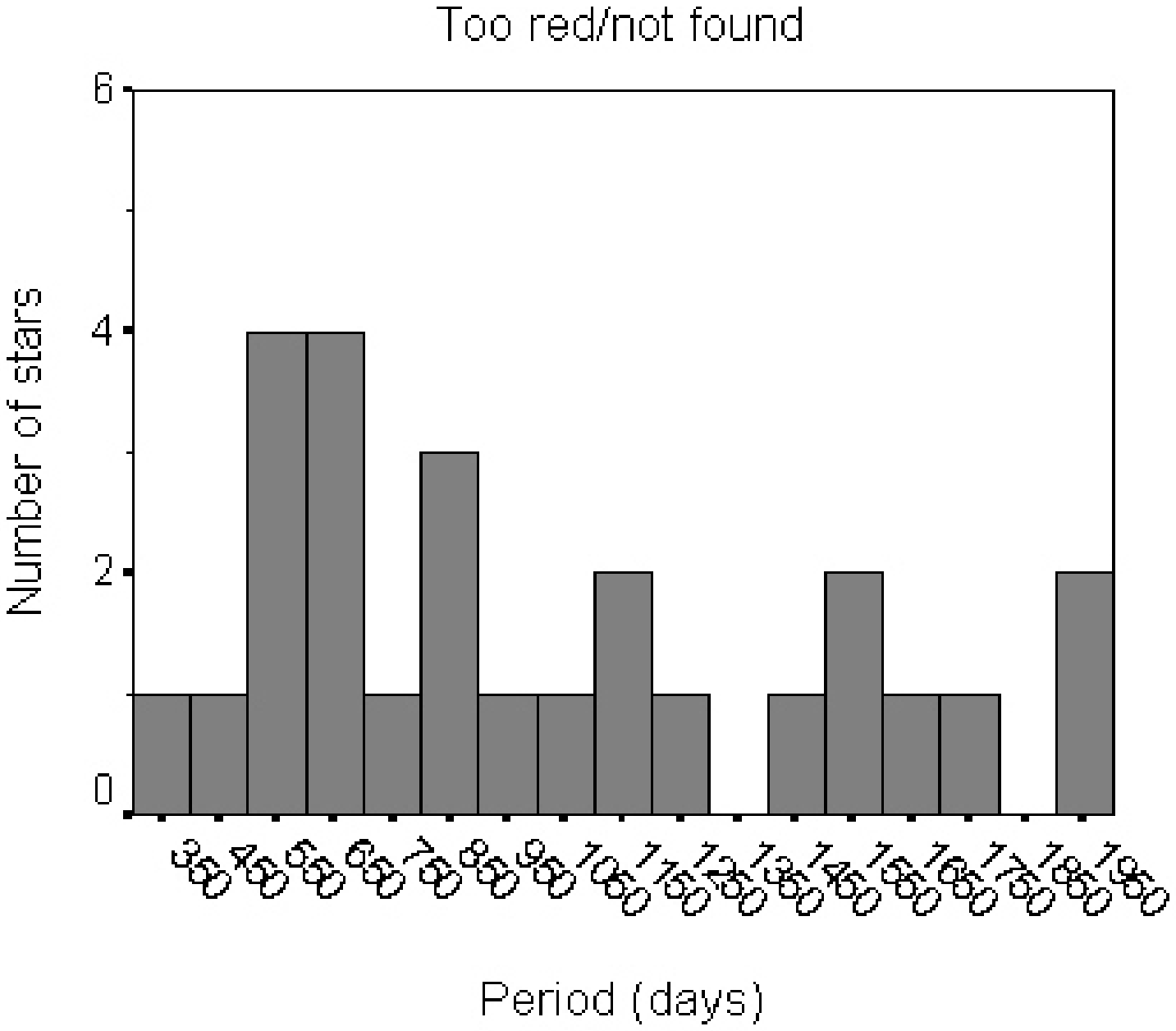}
\caption{Distribution of variability periods for the Li non-detected
(left panel), Li detected (middle panel) and too red/not found (right panel)
sources.}
\end{figure*}

\begin{figure*}
\centering
\includegraphics[width=5.9cm]{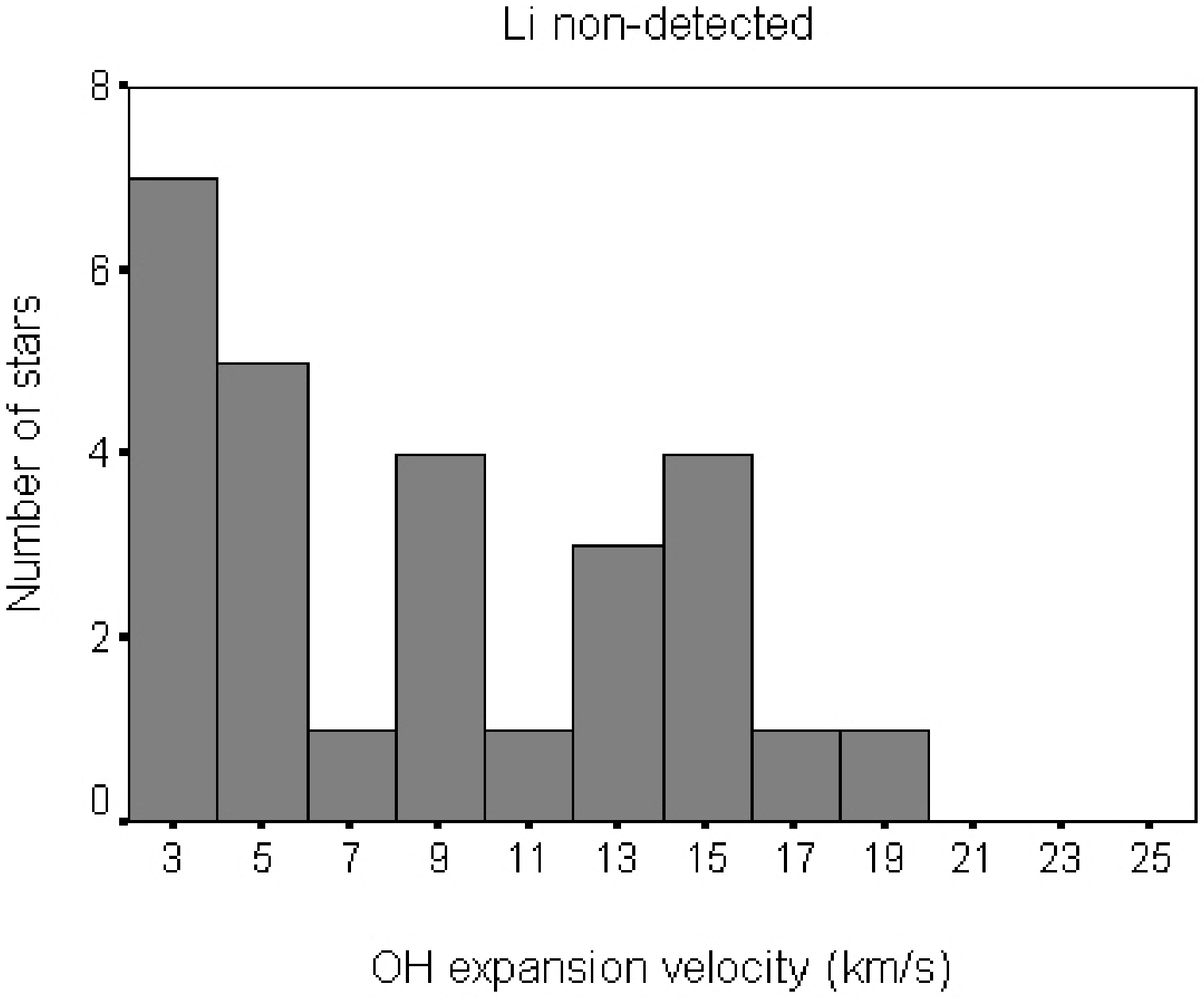}%
\includegraphics[width=5.9cm]{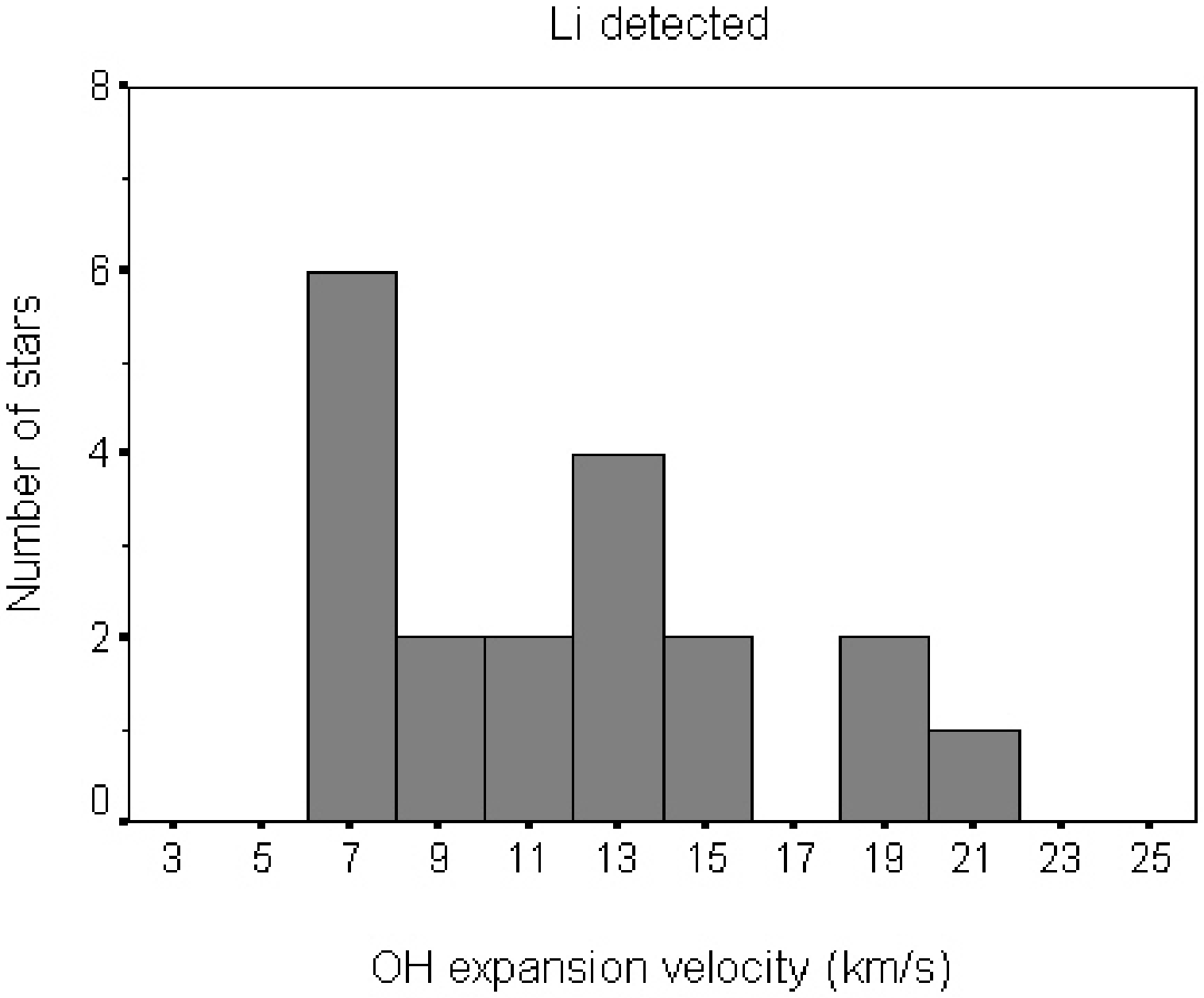}%
\includegraphics[width=5.9cm]{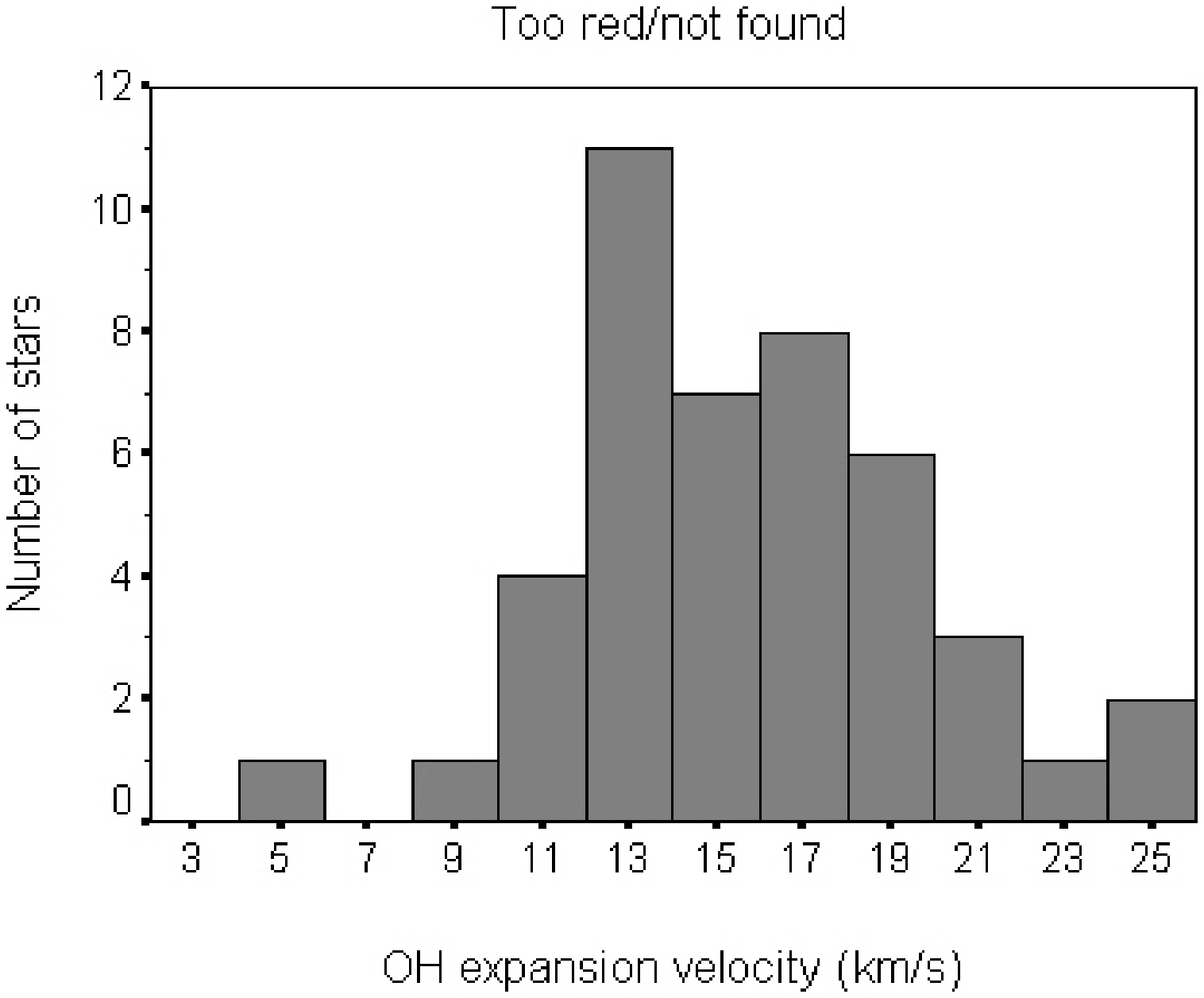}
\caption{Distribution of OH expansion velocities for the
Li undetected (left panel), Li detected (middle panel) and too red/not found
(right panel) sources.}
\end{figure*}

\begin{table*}
\centering
\caption[]{Distribution of sources among the different subgroups identified in 
the sample as a function of their OH expansion velocity and variability 
properties.}
\vspace{4.mm}
\footnotesize
\begin{tabular}{ccccc}
\hline\hline
Period (days)& Observed& Li undetected& Li detected& Too red/not found\\
\hline
$<$400& 12& 10 (84\%) & 1 (8\%) & 1 (8\%)\\
400$-$700& 26& 4 (15\%) & 12 (46\%)& 10 (39\%)\\
$>$700& 17& 1 (6\%) & 1 (6\%) & 15 (88\%)\\
\hline
\end{tabular}
\centering
\footnotesize

\vspace{0.5cm}
\begin{tabular}{ccccc}
\hline\hline
$v_{\rm exp}(OH)$ (km s$^{-1}$)& Observed& Li undetected& Li detected& Too red/not found\\
\hline
$<$6 & 13& 12 (92\%)& 0 (0\%)& 1 (8\%)\\
6$-$12& 25& 6 (24\%)& 11 (44\%)& 8 (32\%)\\
$>$12& 53& 10 (19\%)& 8 (15\%)& 35 (66\%)\\
\hline
\end{tabular}
\end{table*}

 In Table 12 we can see that most of the AGB stars with the  lower OH expansion
velocities ($v_{\rm exp}$(OH) $<$ 6 km s$^{-1}$) are among those undetected in
the Li I line (92\%) while 44\% of stars with OH  expansion velocities between
6 and 12 km s$^{-1}$ (but still observable in  the optical) are detected as
Li-strong sources. However, there are also stars  with non-detection of lithium
(24\%) and stars too red or not found (32\%)  within this group. Finally, a
majority of the stars in the group  of heavily obscured sources (too red or not
found) shows OH expansion  velocities higher than 12 km s$^{-1}$ (66\%).

 Similar statistics are obtained for the variability period. Sources with
relatively short periods, i.e.\ below 400 days, are predominantly non-Li
detections (84\%), while almost half of the stars showing periods between 400
and 700 days are Li detected (46\%). Again, the same observational problems
arise when we try to analyse the stars with the longer periods ($P >$ 700
days), since most of them are completely obscured in the optical (88\%). 

\subsection{Period and OH expansion velocity versus HBB}

In Figure 17 we can see that there is actually a clear correlation between the
Li abundance and the variability period in  our target stars. Remarkably, for
periods below 400 days ({log $P$}  $<$ 2.6), only one star out of 12 is Li
detected. In contrast, almost all  stars with $P >$ 500 days ({log $P$}
$>$ 2.7) are Li-rich, the only exception being IRAS 18050$-$2213 (= VX Sgr),
which is a peculiar, very  cool semi-regular pulsating star, not showing a
proper Mira variability in the last decade, which some authors identify as a
massive red supergiant, and not as an AGB star  (Kamohara et al.\ 2005).

A similar correlation, although not so clear, is found when the expansion 
velocities derived from the OH maser measurements are considered  (see Figure
18). In this case we can see that the maximum Li abundance increases with the
OH expansion velocity, suggesting that the most massive AGB stars can
experience a higher lithium enrichment. However, the more interesting result is
the non-detection of Li in any star of the sample with  {$v_{\rm exp}$(OH)}
$<$ 6 km s$^{-1}$. IRAS 18050$-$2213 is again one of the two outliers showing
no Li and a very high expansion velocity,  while the other one is IRAS
13442$-$6109, for still unknown reasons.

At present, it is generally accepted that lithium production in massive O-rich
AGB stars is due to the activation of hot bottom burning
at the bottom of the convective mantle, which prevents the formation of
luminous C-rich AGB stars (see details in Section 1). If the detection of
lithium is taken as a signature of HBB and the period and  the OH expansion
velocity are accepted as valid mass indicators, our results indicate that there
are no HBB stars among the subgroup of AGB stars in our  sample with periods
lower than 400 days and $v_{\rm exp}$(OH) $<$ 6 km s$^{-1}$,  while most of the AGB
stars with periods beyond 500 days and high expansion  velocities are quite
probably developing HBB.

\begin{figure}
\resizebox{\hsize}{!}{\includegraphics[width=8.5cm,height=9.5cm,angle=-90]{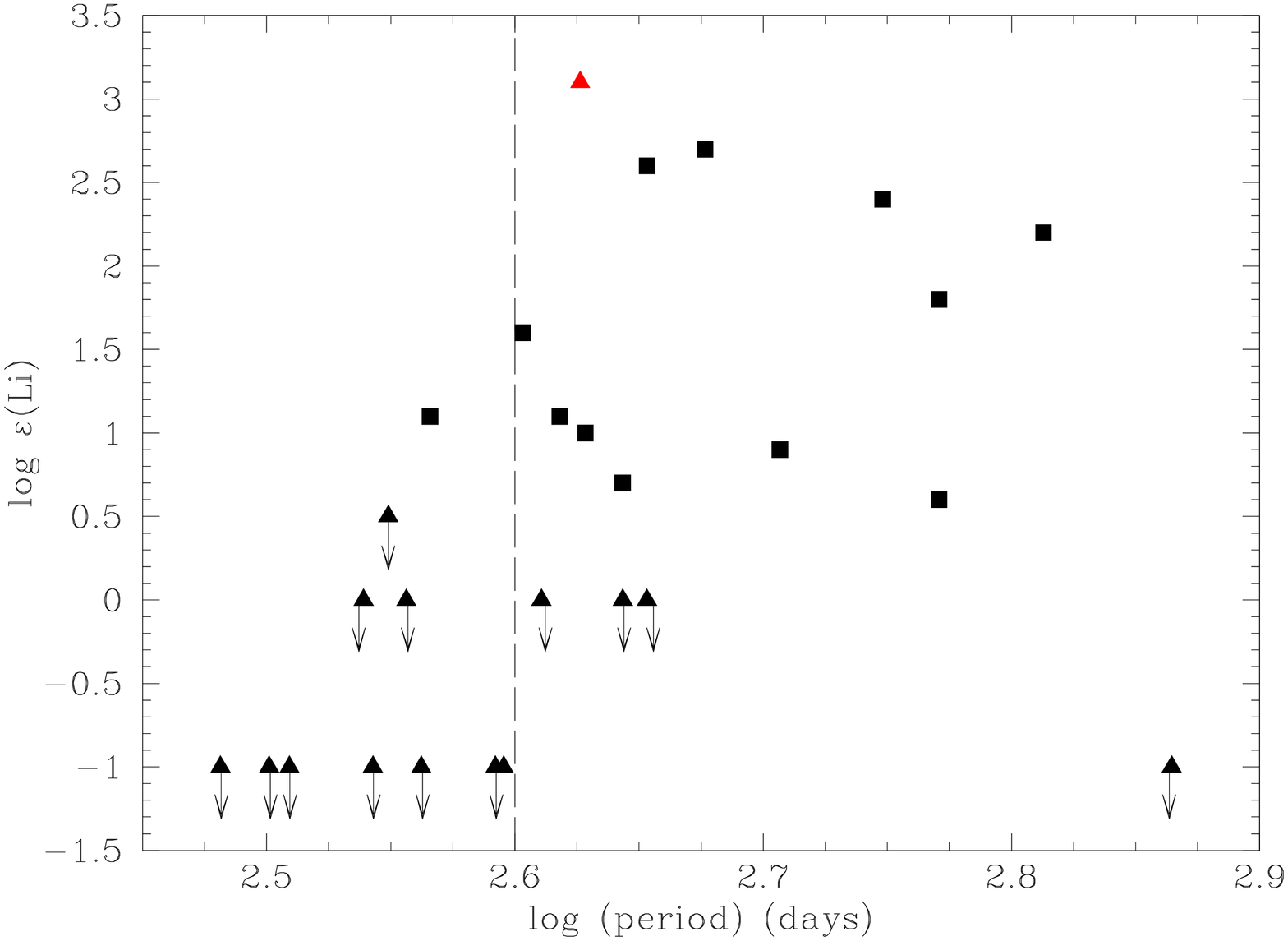}}
\caption{Observed Li abundance vs variability period. 
Upper limits to the Li abundance are
shown with black triangles and marked with arrows. Abundance values derived
from  Li I lines which are resolved in two components (circumstellar and
stellar) are shown with red triangles and correspond to the photospheric
abundance needed to fit the stellar component. The dashed vertical line 
corresponds to a variability period of 400 days (see text).}
\end{figure}

\begin{figure}
\resizebox{\hsize}{!}{\includegraphics[width=8.5cm,height=9.5cm,angle=-90]{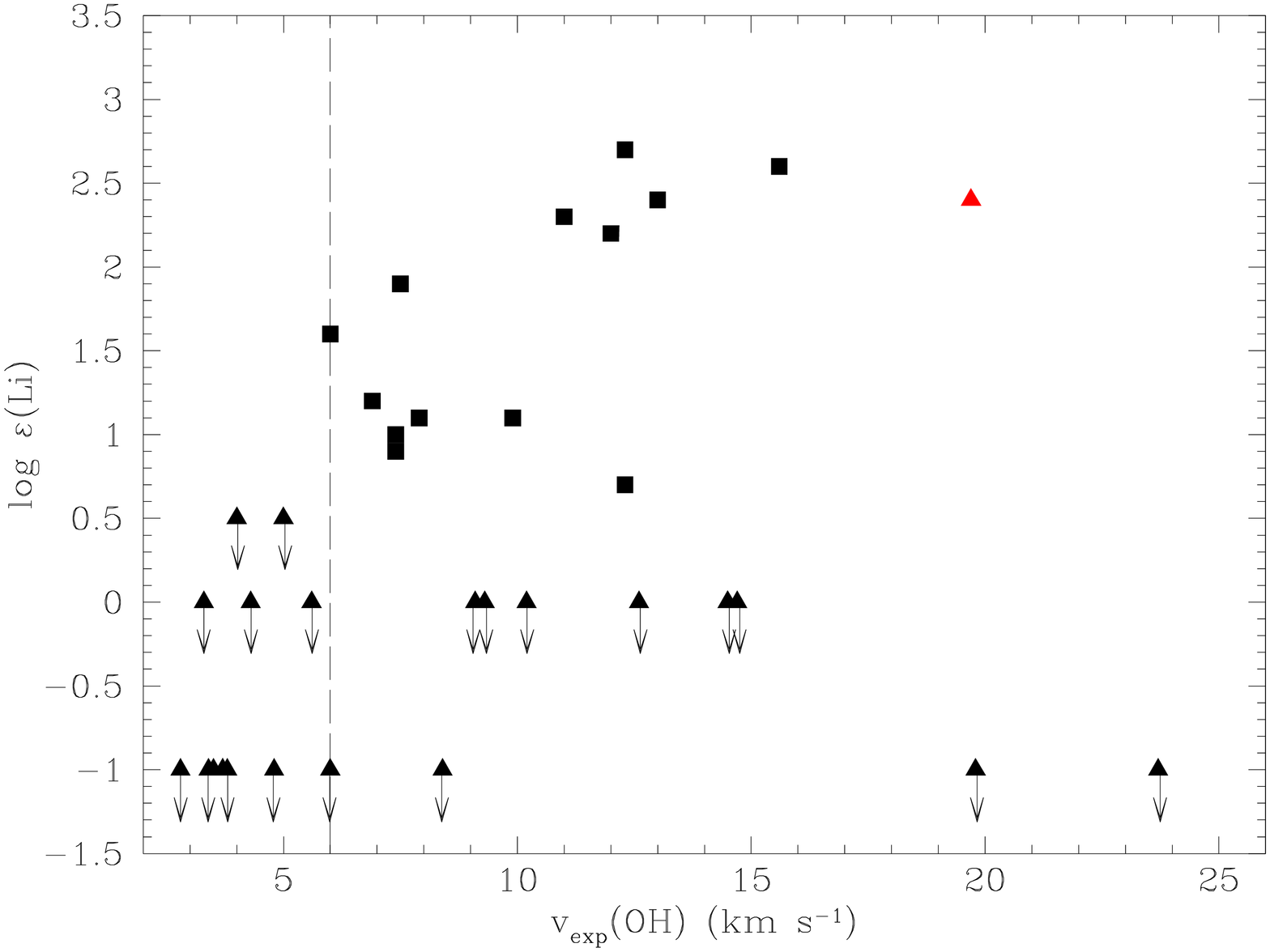}}
\caption{Observed Li abundance vs OH expansion velocity. Upper limits to the
Li abundance are shown with black triangles and marked with arrows. The
abundance values derived from  Li I lines which are resolved into two components
(circumstellar and stellar) are shown with red triangles and correspond to the
photospheric abundance needed to fit the stellar component. The dashed vertical
line corresponds to an OH expansion velocity of 6 km s$^{-1}$ (see
text).}
\end{figure}

\subsection{Lithium production and HBB model predictions}
Modelling of HBB started thirty years ago with envelope models including
non-instantaneous mixing coupled to the nuclear evolution (Sackmann, Smith \&
Despain 1974). More recent models (Sackmann \& Boothroyd 1992; Mazzitelli,
D'Antona \& Ventura 1999, hereafter SB92 and MDV99, respectively) seem to work
well in reproducing the lithium production observed in massive O-rich AGB stars in
the MCs. Although the precise determination of the lithium production depends
on the input physics (stellar mass, metallicity, mass loss rate, overshooting,
etc.), the strongest dependence is with the convection model assumed. SB92
models predict lithium production in intermediate mass stars assuming the
so-called  ``mixing length theory'' (MLT) framework for convection. But these
models require a fine tuning of the mixing length parameter $\alpha$. MDV99
explored lithium production in intermediate mass stars according to a more
modern treatment of turbulent convection, known as the ``Full Spectrum of
Turbulence'' (FST) model (D'Antona \& Mazzitelli 1996, and references therein).
Actually, MDV99 models are able to reproduce the strong Li enhancements 
observed in massive O-rich LMC AGB stars in a satisfactory way and  provide a
theoretical explanation to the existence  of a few sources in the LMC with
extreme luminosities of up to $M_{\rm bol}$ $=$ $-$7.3 and  $-$7.6, which are known
to be long period, obscured AGB stars (Ventura, D'Antona \&  Mazzitelli 2000).
However, these models have not yet been tested against their Galactic analogues.

 For solar metallicity, MDV99 predict lithium production as a consequence of 
the activation of HBB for stellar masses $\gtrsim$ 4.5 $M_\odot$ without core
overshooting and for $M \gtrsim  4\ M_\odot$ including core overshooting. The
amount of lithium produced during the AGB does not depend on the assumed 
initial lithium abundance as the star structure loses all memory of  the
previous history of lithium at the beginning of the AGB phase. As an example,
according to this model the lithium surface abundance drops to {log
$\varepsilon$(Li)} $\sim$ 0.3 for a 6 $M_\odot$ star after the second
dredge-up, at the beginning of the AGB phase. If overshooting from below the
convective envelope is allowed, an important decrease in the number of thermal
pulses is predicted by the model and lithium production takes place before the
first thermal pulse. The variation with time of the lithium abundance, total
luminosity and temperature at the base of the convective envelope ($T_{\rm bce}$)
for different masses is shown in Figure 15 of MDV99. A stronger
Li-overabundance, faster increase of the luminosity at the beginning of the AGB
phase and higher $T_{\rm bce}$ are predicted when the mass of the star is
increased. 

Mass loss rates and metallicities play also an important role according to
MDV99. The minimum mass needed for the ignition of HBB decreases with 
decreasing mass loss rates. In addition, a low mass loss rate leads to a longer
run of  thermal pulses, and to a lithium abundance which remains large for the
entire thermal pulsing AGB (TP-AGB, hereafter) lifetime. In contrast, for high
mass loss rates, the HBB process is stopped only after a few pulses ($\sim$5
for a 6 $M_\odot$ star). Models with strong mass loss show oscillations of the
Li abundance by orders of magnitude on short timescales of $\sim$10$^{4}$ years
(even within a single thermal pulse!). The temperature at the bottom of the
convective envelope, $T_{\rm bce}$, varies so much that the lithium production is
temporarily stopped while the lithium already produced has already been diluted
by convection.  When $T_{\rm bce}$ increases, the lithium abundance increases
again. The consequence is that, at least for 20\% of the time, there is
negligible lithium in the envelope (at a level of approximately {log
$\varepsilon$(Li)} $\lesssim$ 1.0). This implies that the distribution of Li
abundances derived from the observations can only be analysed in a statistical
way. 

On the other hand, the activation of HBB takes place at a lower mass limit of
only 3.0--3.7 $M_\odot$ (depending on the mass loss rate considered) at the
metallicity of the LMC (Ventura, D'Antona \& Mazzitelli 2000). Low metallicity
models predict larger values of $T_{\rm bce}$ and higher luminosities, and as a
consequence, higher lithium production for a given stellar mass. In addition,
this accelerates Li-production and a maximum value of {log
$\varepsilon$(Li)} $\sim$ 4.3 is reached in a shorter timescale with respect to
the solar metallicity case. The variation with time of the surface lithium
abundance, stellar luminosity and $T_{\rm bce}$, computed for a 6 $M_\odot$ model
and different metallicities is shown in Figure 10 of MDV99. Lower metallicity
models esentially produce lithium faster due to the larger temperature at the
base of the convective envelope, but show high Li abundances for a shorter
time.

\subsection{Theory versus observations}
Globally considered, our results indicate that below a certain value of the
period and the OH expansion velocity our stars do not show any Li enhancement.
However, not all the stars above these values show  detectable Li. The strong
IR excesses associated with most of the stars in our sample suggest  that they
are actually at the end of their evolution as AGB stars.  Surprisingly, none
of the stars in our sample shows a strong s-process enhancement (e.g.\ from Zr,
Y, Sr) as  would have been expected if they were stars that have experienced
a significant third dredge-up during their AGB evolution. In the case of the
more massive stars in our sample one could argue that the mass loss is so
strong at the high metallicity of our Galaxy (compared to the MC metallicity)
that their AGB lifetime may become shortened significantly, so that they will
end this phase experiencing too few thermal pulses before they completely lose
their envelope. However, the lack of s-process enhancement is also observed in
the group of stars in which no lithium has been detected.

In order to interpret these results, several scenarios can be proposed:
\begin{enumerate}
\item[i)] all the stars in the sample are HBB stars, and thus, considerably
massive ($M\gtrsim  4\ M_\odot$). The difference between the Li detected and
the Li undetected ones would be just that the undetected ones are in a very
early stage as AGB stars. This hypothesis would explain the absence of
s-process enhancement (e.g.\ Zr, Y, Sr) in all the stars in the sample,  but it
would not be consistent with the non-detection of lithium in the group of stars
with the shorter periods and lower OH expansion velocities. HBB models predict
a considerable Li enhancement just after the arrival at the AGB. Statistically,
there should be no difference in the behaviour of the Li abundances for any
star in the sample, and this is not observed.

\item[ii)] the sample is composed by a combination of low mass ($M\lesssim 
1.5\ M_\odot$) and high mass ($M\gtrsim  4\ M_\odot$) stars. In this scenario,
the optically bright O-rich AGB stars in our sample with the shorter periods
and lower OH expansion velocities, undetected in Li, would belong to the
population of low mass stars  in the Galaxy that do not experience HBB and
evolve all the way along the AGB as O-rich stars. However, these stars are
expected to evolve very slowly and develop a large number of thermal pulses
during the AGB. At the end of their AGB evolution they are expected to become
strongly enhanced in s-process elements as a consequence of the dredge-up of
processed material induced by the many thermal pulses experienced. But the Li
undetected stars in our sample do not show this enhancement in s-process
elements. The only way to reconcile this result with their status as low mass
stars would imply  assuming that they are at a very early stage as AGB stars,
which again seems to be incompatible with the strong IR excess detected by
IRAS. Such a strong excess is only expected in low mass stars at the very end of
the AGB. Alternatively, they could experience less thermal pulses than
predicted by the models or the dredge-up process could be very inefficient. But
this is not what we observe in other low  and intermediate mass stars in the
Galaxy, such as the S-stars and many C-rich stars, which show the strong s-process
enhancement that is not observed in our stars. Therefore, we must conclude
that the hypothesis of these Li undetected stars being identified as low mass
AGB stars seems to be implausible.

\item[iii)] all the stars in the sample are intermediate mass stars
($M\gtrsim 3\ M_\odot$) but only a fraction of them, the more massive ones 
($M\gtrsim 4\ M_\odot$), experience HBB. This scenario would explain why the
sources with lower periods and OH expansion velocities in our sample are
undetected in the Li I line, while at the same time, they do not show any
s-process enhancement. These stars, although slightly below the minimum mass
needed to develop HBB, would also experience a very strong mass loss rate and
only a few thermal pulses ($\sim$5--10) before the early termination of the
TP-AGB as a consequence of the strong mass loss. The initial envelope of these
stars would also be quite massive and thus a slow change of the C/O ratio  is
expected because of the huge dilution of the dredged up material. But how can
we explain the large spread of Li abundances observed in the  group of sources
in our sample with longer periods and larger OH expansion velocities? These are all
supposed to be HBB stars. And HBB models predict a large fluctuation in the
Li abundance as the star evolves along the TP-AGB (see figure 15 of MDV99
again) between two consecutive pulses. Actually,  the timescale of the Li
production phase in the TP-AGB is predicted to be of the order or slightly
smaller than the timescale between pulses ($\sim$10$^{4}$ years for a massive
star with $M\gtrsim\ 4 M_\odot$). In addition, as the AGB evolution proceeds,
the surface Li abundances decrease again very rapidly as a consequence of the
$^{3}$He becoming almost completely burned in the envelope and thus  the Li
production ceases (e.g.\ MDV99; Forestini \& Charbonnel 1997). This would be
consistent with the detection of strong Li abundances in $\sim$50\% of the
stars in our sample for which we suspect that HBB is active. In this
scenario, the strongly obscured sources for  which we were not able to obtain
an optical spectrum could also be interpreted as a population of high mass
stars developing HBB with even more extreme observational properties. They are
sources with longer periods ($>$ 700 days) and larger OH expansion velocities
($>$ 12 km s$^{-1}$) which may have  developed optically thick circumstellar
envelopes as a consequence of the huge mass loss experienced in the AGB. These
stars must also be detectable as Li-rich at least for some time during their
AGB evolution but, unfortunately, they are so strongly reddened in the optical
range that their study is only possible in the infrared, where no strong Li
lines are present. We propose that the HBB nature of these obscured stars can
be elucidated by determining their \Crat\ ratios, which can also be used as an
HBB indicator, through near infrared spectroscopy. 
\end{enumerate}

\subsection{Comparison with the O-rich AGB stars in the Magellanic Clouds}
The situation is quite different for the AGB stars in the MCs. In contrast with
their Galactic analogues, the more luminous AGB stars in the MCs (with $-$7
$\le$ $M_{\rm bol}$ $\le$ $-$6) are O-rich stars showing s-process element
enhancements. In addition, a higher proportion of them ($\sim$80\% compared to
$\sim$50\% in our Galaxy)  also shows  lithium overabundances. The enrichment in
lithium indicates that they are also HBB stars, but why are these stars also
enriched in s-process elements? 

The answer to this question must be related to the different metallicity of the
stars in the MCs with respect to our Galaxy. Actually, theoretical models
predict a higher efficiency of the dredge-up in low metallicity  atmospheres
(e.g.\ Herwig 2004) with respect to those with solar metallicity  (e.g.\ Lugaro
et al.\ 2003). 

Moreover, there is observational evidence that lower metallicity environments
are also less favourable to dust production, as  is suggested by the smaller
number of heavily obscured AGB stars in the MCs (Trams et al.\ 1999; Groenewegen
et al.\ 2000), compared to our Galaxy. This is supported by the lower
dust-to-gas ratios derived by van Loon (2000) in the few obscured AGB stars in
the MCs for which this analysis has been made.

If mass loss is driven by radiation pressure on the dust grains, it might be
less efficient with decreasing metallicity (Willson 2000). In that case, longer
AGB lifetimes would be expected, which could increase the chance of
nuclear-processed material  reaching the stellar surface. This would also
explain the larger proportion of luminous C-rich stars  (up to
$M_{\rm bol}$ $\sim$ $-$6) observed in the MCs (Plez, Smith \& Lambert 1993; Smith \&
Lambert 1989, 1990a; Smith et al.\ 1995) with respect to the Galaxy. The slow
evolution predicted for AGB stars in the MCs as a consequence of the less 
efficient mass loss leaves time for more thermal pulses to occur during the 
AGB lifetime and, therefore, a more effective dredge-up of s-process elements
to the surface can be expected before the envelope is completely gone at the
end of the AGB. This would explain why even the more massive stars in the MCs
show a strong s-process enrichment in contrast to their Galactic counterparts.
In our Galaxy the only AGB stars showing a similar overabundance in s-process
elements seem to be the result of the evolution of low  to intermediate mass
stars ($M\lesssim$ 1.5--2.0 $M_\odot$), while no, or very little, s-process
enhancement is observed in Galactic AGB stars with higher main sequence
masses. 

In addition, the lower critical mass (which also decreases with decreasing mass
loss rate, see Section 5.4) needed to develop HBB ($M\gtrsim$ 3 $M_\odot$ at
the metallicity of the LMC, compared to the $\sim$4 $M_\odot$ limit in our
Galaxy) favours the simultaneous detection of s-process elements and Li
enrichment in a larger number of AGB stars in the MCs. In our Galaxy, the
 Li-rich sample of AGB stars is restricted to the fraction of stars with main
sequence masses  $M \gtrsim$ 4 $M_\odot$. In contrast to their MC counterparts,
these stars evolve so rapidly (because of the strong mass loss) that there is
no time for a significant enhancement in s-process elements, as the results
here presented seems to demonstrate.

Obviously, more refined s-process calculations taking into account the effects
of HBB and mass loss,  using a large grid of stellar evolutionary models with
different masses and metallicities are strongly encouraged.

\subsection{Evolutionary connections}
The well known existence of at least two different chemical branches of AGB 
stars (C-rich and O-rich) and the dependence of this chemical segregation on 
the main sequence mass and on the environmental conditions in which these stars
evolve (such as the metallicity) is a long debated issue in stellar evolution.
Moreover, the connection with the chemical composition observed in Galactic
post-AGB stars and/or well evolved PNe (which are expected to be the result of
the evolution of AGB stars with a wide variety of masses) has never been
addressed in detail from an overall perspective.

Stars classified as post-AGB in the literature are mainly low mass stars,
according to their wide Galactic distribution. They are thought to  represent
the fraction of stars which evolve so slowly that they are still  visible for
some time as stars with intermediate spectral types before  they become PNe.
Many of these stars are C-rich and show strong s-process element enhancements
[s-process/Fe] $\sim$ $+$1.5 (van Winckel \& Reyniers 2000; Reddy, Baker \&
Hrivnak 1999; Klochkova et al.\ 1999), in agreement with what we would expect if
they were the final product of the evolution of low mass AGB stars. However,
other post-AGB stars do not show the characteristic signatures of the third
dredge-up. In the non-enriched sources the s-process element abundances can be
as low as [Zr/Fe] $\lesssim$ 0.2 (Luck, Bond \& Lambert 1990;  van Winckel
1997). These objects might be stars with even lower masses ($M\sim$ 0.8--1.0
$M_\odot$) which have evolved off the AGB before the third dredge-up could
occur (Marigo, Girardi \& Bressan 1999). Some of these sources are
high galactic latitude, hot  B-type post-AGB stars (e.g.\ Conlon et al.\ 1993; 
Mooney et al.\ 2001) that show strong carbon deficiencies. In the more extreme
cases they will probably never become PNe. 

On the other hand, and as we have shown in this article, the more massive AGB
stars may also show a similar lack  of s-process elements, but in this case as
a consequence of the rapid  evolution induced by the strong mass loss. These
stars are more  difficult to observe in the optical as they are expected to
evolve as heavily obscured  sources  from the TP-AGB to the PN stage but may be
the progenitors of most Galactic PNe.

Actually, a similar chemical segregation is also observed in Galactic PNe. 
Classically, they are classified as type I, II or III as a function of their
chemical abundance properties (Peimbert 1978). They are expected to cover a
wide range of progenitor masses and thus show different chemical signatures
as a consequence of their previous passage along the AGB. 

In particular, the higher mass fraction of stars (with $M >$ 4 $M_\odot$) 
identified in this paper as stars developing HBB and showing a strong lithium
enhancement are expected to arrive at the PN stage still as heavily obscured
stars. These stars would evolve into N-rich type I PNe, which are
characterized by their large He abundances (He/H $>$ 0.10) and N/O ratios
(Manchado 2003, 2004; Mampaso 2004, and references therein). They are thought
to represent the fraction of the more massive PNe in the Galaxy, as  is
confirmed by their strong concentration at low galactic latitudes. Their strong
N overabundances would be consistent with their identification as PNe resulting
from the evolution of HBB AGB stars.

\section{Conclusions}
As a result of a high resolution spectroscopic survey carried out in the
optical for a large sample (102) of massive Galactic O-rich AGB stars, we have
detected the Li I resonance line at 6708 \AA~in 25 stars, while 32 stars did
not display this Li I line. The rest of the stars observed (45 stars) were
heavily obscured by their thick circumstellar envelopes and, as such, they 
were too red or not found at optical wavelengths. 

We have classified the sample into three subgroups on the basis of their OH
expansion velocity and variability period. From their relative distribution in
the IRAS two--colour diagram [12] $-$ [25] vs [25] $-$ [60] and Galactic distribution,
we conclude that they must represent populations of Galactic O-rich AGB stars
with different progenitor masses.

Combining MARCS model atmospheres and synthetic spectroscopy with extensive
line lists we have been able to derive the fundamental stellar parameters
($T_{\rm eff}$, log $g$, C/O, $\xi$, etc.) and obtain the Li and Zr
abundances in those stars for which an optical spectrum was obtained. Our
chemical abundance analysis shows that half of stars show Li overabundances in
the range between {log $\varepsilon$(Li)} $\sim$ $+$0.5 and $+$3.0 dex.
This is interpreted as a signature of the activation of HBB, confirming that
they are actually massive AGB stars ($M >$ 4 $M_\odot$). However, these stars
do not show any zirconium enhancement (taken as  representative for the
s-process element enrichment), indicating that they behave differently from the
more massive (and luminous) AGB stars in the MCs. 

Assuming that the variability period and the OH expansion velocity can be taken
as distance-independent mass indicators and by comparing our results with the
theoretical predictions of HBB and nucleosynthesis models we conclude that the
O-rich AGB stars in our sample are all considerably massive ($M >$ 3 $M_\odot$)
even those non-detected in lithium, but only the more massive ones ($M >$ 4
$M_\odot$; with periods longer than 400 days and {$v_{\rm exp}$(OH)} $>$ 6
km s$^{-1}$) experience HBB. The lack of lithium enrichment found in some of 
the sources  thought to belong to the group of more massive AGB stars in the
sample is explained as the consequence of the timescale of the Li production
phase being of the order of or slightly smaller than the interpulse time
($\sim$10$^{4}$ years) or, alternatively, by the $^{3}$He exhaustion in the
envelope  interrupting the Li production, in agreement with the predictions
made by the models. 

The group of stars displaying the most extreme observational properties (with
periods sometimes longer than 700 days and {$v_{\rm exp}$(OH)} $>$ 12 km
s$^{-1}$) are believed to represent the more massive AGB stars in our Galaxy.
Unfortunately, these stars are strongly obscured by their thick circumstellar
envelopes and we could not carry out any type of analysis in the optical range.
This very high mass population should also experience HBB and show strong Li
enhancement but no s-process element overabundances. We propose that the HBB
status of these obscured stars can be determined by measuring their \Crat\
ratios in the near-infrared, which can also be used as a HBB indicator.

Our results suggest that the dramatically different abundance pattern found in
AGB stars belonging to the MCs and to our Galaxy can be explained in terms of
the different metallicity conditions under which these stars evolved. This 
constitutes strong observational evidence that the chemical evolution of 
massive AGB stars is strongly modulated by metallicity.

Globally considered, our results can also be used to establish evolutionary
connections between TP-AGB stars, post-AGB stars and PNe. In particular, we
suggest that the HBB AGB stars identified in our sample must be the precursors
of N-rich type I PNe.

\begin{acknowledgements}
DAGH is grateful to J.I. Gonz\'alez-Hern\'andez from the Instituto de
Astrof\'{\i}sica de Canarias for the adaptation of his $\chi^2$ code used in
this work. AM and PGL acknowledge support from grant \emph{AYA 2004$-$3136}
and \emph{AYA 2003$-$9499} from the Spanish Ministerio de Educaci\'on y 
Ciencia.   
\end{acknowledgements}

\end{document}